\documentclass[a4paper,fleqn,usenatbib]{mnras}

\usepackage[T1]{fontenc}
\usepackage{ae,aecompl}

\usepackage{graphicx}	
\usepackage{amsmath}	
\usepackage{amssymb}	
\usepackage{float}
\usepackage{tabularx}

\title[Jet Power]{Jets in Magnetically Arrested Hot Accretion Flows: Geometry, Power and Black Hole Spindown }

\author[Narayan et al.]{
Ramesh Narayan,$^{1,2}$\thanks{E-mail: rnarayan@cfa.harvard.edu}
Andrew Chael,$^{3,4}$
Koushik Chatterjee,$^{1,2}$
Angelo Ricarte,$^{1,2}$ 
\newauthor Brandon Curd$^{1,2}$
\\
$^{1}$ Harvard-Smithsonian Center for Astrophysics, 60 Garden Street, Cambridge, MA 02138, USA
\\
$^{2}$ Black Hole Initiative at Harvard University, 20 Garden Street, Cambridge, MA 02138, USA
\\
$^{3}$ Princeton Center for Theoretical Science, Princeton University, Princeton, NJ 08544, USA
\\
$^{4}$ NASA Hubble Fellowship Program, Einstein Fellow
}

\date{Accepted 2022 January 26. Received 2022 January 26; in original form 2021 August 27}

\pubyear{2022}

\begin{document}
\label{firstpage}
\pagerange{\pageref{firstpage}--\pageref{lastpage}}
\maketitle

\begin{abstract}
We present the results of nine simulations of radiatively-inefficient magnetically arrested disks (MADs) across different values of the black hole spin parameter $a_*$: $-0.9$, $-0.7$, $-0.5$, $-0.3$, 0, 0.3, 0.5, 0.7, and 0.9.  Each simulation was run up to $t \gtrsim 100,000\,GM/c^3$ to ensure disk inflow equilibrium out to large radii. We find that the saturated magnetic flux level, and consequently also jet power, of MAD disks depends strongly on the black hole spin, confirming previous results.  Prograde disks saturate at a much higher relative magnetic flux and have more powerful jets than their retrograde counterparts. MADs with spinning black holes naturally launch jets with generalized parabolic profiles whose widths vary as a power of distance from the black hole. For distances up to $100GM/c^2$, the power-law index is $k \approx 0.27-0.42$. There is a strong correlation between the disk-jet geometry and the dimensionless magnetic flux, resulting in prograde systems displaying thinner equatorial accretion flows near the black hole and wider jets, compared to retrograde systems. Prograde and retrograde MADs also exhibit different trends in disk variability: accretion rate variability increases with increasing spin for $a_*>0$ and remains almost constant for $a_*\lesssim 0$, while magnetic flux variability shows the opposite trend. Jets in the MAD state remove more angular momentum from black holes than is accreted, effectively spinning down the black hole. If powerful jets from MAD systems in Nature are persistent, this loss of angular momentum will notably reduce the black hole spin over cosmic time.  
\end{abstract}




\section{Introduction} \label{sec:intro}

Hot accretion flows are common in astrophysical black holes (BHs) such as those found in low-luminosity active galactic nuclei (AGNs) and hard-state BH X-ray binaries \citep[see][for a review]{Yuan2014}. Many of these BH systems exhibit relativistic jets \citep[e.g.,][]{PaperI,Fender2001}. Understanding how these jets are powered is important, both because the underlying physics is intrinsically of interest, and because AGN jets often interact with galactic gas and inject energy into the interstellar medium, a process thought to be pivotal to AGN feedback \citep[e.g.,][and references therein]{Harrison2018}. 

Over the past two decades, general relativistic magnetohydrodynamic (GRMHD) simulations have become a popular tool to model hot accretion flows (previously called advection-dominated accretion flows, cf. \citealt{Narayan1994, Narayan1995}, or radiatively inefficient accretion flows). The GRMHD equations account for the dynamical evolution of magnetized plasma in the framework of general relativity, both for spinning and non-spinning BHs, and have been used extensively to predict observables, particularly in jetted BH systems. Simulations have shown that one can produce relativistic jets naturally without any substantial tuning of the initial conditions \citep[e.g.][]{McKinney2004,DeVilliers2005}. As gas spirals in toward the BH, poloidal magnetic field loops are dragged in with the disk gas and are twisted by the BH's frame-dragging effect, enabling an outward pressure that launches a relativistic jet. Since frame-dragging is associated with BH rotation, one expects the BH spin to play a key role in determining the power in the jet. 

A key development in accretion theory was the recognition of the importance of magnetically-dominated accretion flows. Using MHD simulations, \citet[][see also \citealt{Igumenshchev2008}]{Igumenshchev2003} found that, given the right initial conditions, magnetic fields can become dynamically important in BH accretion flows, to the extent that they impede the inward motion of gas and create a ``magnetically arrested disk'' (MAD, \citealt{Narayan2003}, see also \citealt{Bisnovatyi1974,Bisnovatyi1976}). Hot accretion flows in the MAD regime can launch powerful jets, with power at times exceeding the accretion energy at the event horizon. In a pioneering study, \citet{Tchekhovskoy2011} demonstrated that a three-dimensional (3D) GRMHD simulation of a BH with an extreme spin, $a_*\equiv a/M =0.99$ ($M$ is the BH mass), and accreting in the MAD state, produced a jet with power $P_{\rm jet} \approx 1.4\dot{M}_0c^2$, where $\dot{M}_0$ is the mass accretion rate. Since the jet in this simulation carried away more energy than the entire rest mass energy of the accreted gas, it could not be powered purely by accretion. The only explanation is that the jet extracts rotational energy from the BH via the Blandford-Znajek \citep[BZ;][]{Blandford1977} process, a magnetic analogue of the Penrose \citep[][]{Penrose69} process \citep[see][]{Lasota2014}.

While initially it appeared that the MAD state may require special initial conditions, e.g., a strong coherent vertical field, it has become increasingly clear that MAD configurations arise even under less extreme conditions. In important work, \citet{Liska2020} showed that a simulation initialized with a pure toroidal field, self-consistently generated poloidal fields and reached the MAD state after it was run  with sufficiently high spatial resolution and for a long enough time. Since the durations of even ``long'' simulations are a tiny fraction of actual accretion time scales in Nature, this suggests that most hot accretion flows in the universe might be in the MAD state. For example, \citet{ressler_2020b} naturally obtain a MAD final state with no fine-tuning of initial conditions in their GRMHD simulation of the accretion flow in Sagittarius A$^*$, when they fuel the disk via weakly magnetized stellar winds from distant Wolf-Rayet stars \citep{ressler_2020a}. Independently, high angular resolution polarization observations of M87* by the Event Horizon Telescope indicate that the accretion flow in this system is likely to be in the MAD state \citep{EHTVIII}. Other arguments in favor of MAD configurations in Nature can be found in, e.g., \citet[][]{Zamaninasab2014} and \citet{Nemmen2015}.

Since the source of the energy for jets in MAD systems is the BH spin, one expects the energy efficiency $\eta$ of the jet to be a function of the spin parameter $a_*$. In this paper, we analyze MAD GRMHD simulations that cover a range of spin values, both prograde and retrograde, and we explore how $\eta$ and other parameters of the jet depend on the BH spin.

While a BH accreting from a prograde disk gains angular momentum from the accreting gas, it loses angular momentum in the process of powering a jet. Which process dominates? In an early discussion, \citet{Gammie2004} considered a particular sequence of simulations and concluded that, for their sequence, the BH spins down with time if $a_*>0.94$ and spins up for lower values of $a_*$. However, that result was specific to their initial conditions. Since MAD systems produce especially powerful jets, spindown ought to be strongest in such models. With this expectation in mind, we quantify spinup/spindown for a range of BH spin values, considering both prograde and retrograde disks. These analyses are similar to previous work by \citet{Tchekhovskoy2012} and \citet{Tchekhovskoy2012b}\footnote{and also the ``thinner disc TNM11" class of models in \citet{McKinney2012}}, and are an update and validation of their results. 

In addition, we consider yet other jet and disk properties and study how they vary with BH spin and with the sense of rotation (prograde or retrograde) of the accretion disk. We find substantial differences in the shape of the jet, the radial profiles of some gas parameters in the disk, and the time variability of the mass accretion rate.

We caution that the present work is limited to radiatively inefficient (technically non-radiative, since no cooling is included) hot accretion flows in the MAD state. Hot SANE (``standard and normal evolution" \citealt{Narayan2012}) accretion flows are not covered, nor are thin accretion disks or super-Eddington accretion flows. We also note that the results presented here refer to average properties of systems in steady state. A given system could, at a particular instant, have significant transient deviations.

In Section~\ref{sec:sims}, we describe our numerical code and the initial conditions we use for the simulations. We also discuss common diagnostics for GRMHD simulations, namely the time dependence of the accretion rate and the magnetic flux at the event horizon. In Section~\ref{sec:results}, we discuss our simulation results in several subsections, focusing on the time-averaged behaviour of the horizon magnetic flux, the jet power, the disk and jet structure, and the spindown of the BH. We also study the time variability of relevant quantities. In Section~\ref{sec:discussion}, we discuss the correlation between the MAD magnetic flux saturation level and the disk and jet geometry, and also the effect of spindown on the BH spin evolution. Finally, we summarize our main findings in Section~\ref{sec:summary}.

\section{Simulations}
\label{sec:sims}
\begin{figure}
	\includegraphics[width=8cm]{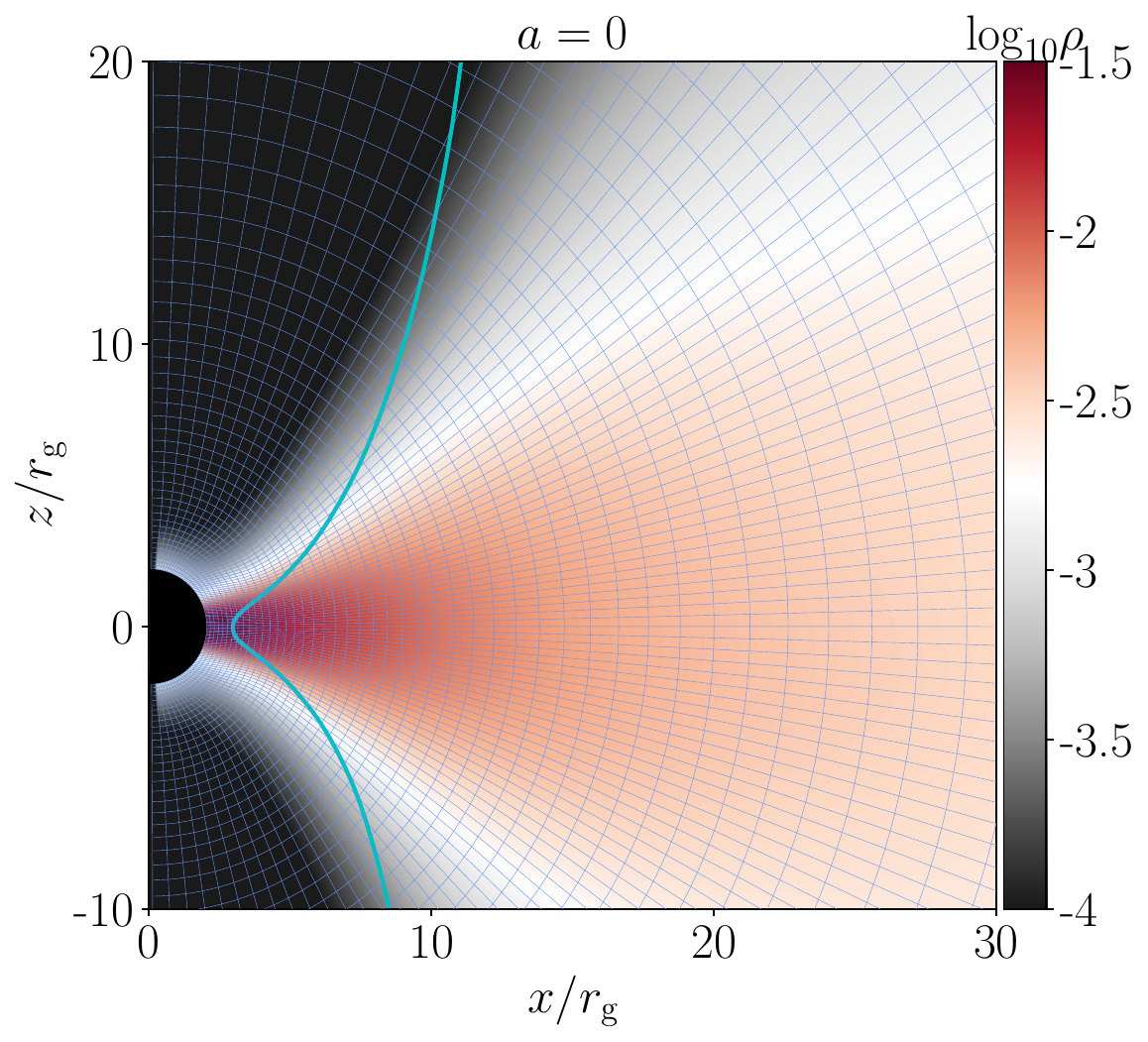}
    \caption{Time- and azimuth-averaged gas density in the poloidal plane of the zero spin $a_*=0$ simulation. The time-averaging was carried out between $50,000$ and $100,000$ $t_{\rm g}$. The blue lines show the simulation grid, which uses a coordinate transformation introduced by \citet{Ressler17} to concentrate resolution in the jet and disk regions near the black hole. For clarity, we show the grid coarse-grained by a factor of two. The cyan contour marks the surface where the magnetization $\sigma_{\rm M}=1.$}
    \label{fig:grid}
\end{figure}

\subsection{The KORAL Code and GRMHD equations}

The simulations described in this paper were run using the GRMHD code KORAL \citep{Sadowski2013, Sadowski2014}.  KORAL is designed to simulate BH accretion and outflow across a wide parameter space, and includes several physical effects that go beyond the assumptions of standard GRMHD. KORAL was initially developed to evolve radiation as well as magnetized gas in accretion flows \citep{Sadowski2013}. It was then extended to evolve separate electron and ion entropies in a two-temperature plasma \citep{KORAL16}, and even to evolve a full distribution of relativistic electrons in addition to the usual thermal population 
\citep{Chael17}. In the present work, since we are interested in radiatively inefficient accretion flows, we only consider standard GRMHD; this version of KORAL has been benchmarked and validated against a number of other GRMHD codes in simulations of both low-magnetic-flux SANE accretion disks \citep{Porth2019} and high-magnetic-flux MAD systems \citep{olivaresinprep}.

KORAL evolves magnetized gas in the Kerr metric.  In standard GRMHD, using gravitational units where $G=c=1$, the energy momentum tensor takes the form: 
\begin{equation}
 T^\mu_{\;\;\nu} = \left(\rho + u + p + b^2\right)u^\mu u_\nu + \left(p + 
\frac{1}{2}b^2\right)\delta^\mu_{\;\;\;\nu} - b^\mu b_\nu.
\label{eq:Tmunu}
\end{equation}
The quantities in $T^\mu_{\;\;\nu}$ evolved by a GRMHD code include the rest mass density $\rho$, fluid internal energy $u$, four-velocity $u^\mu$, and 
the lab frame magnetic field three-vector $B^i$, from which we compute the magnetic four-vector $b^\mu$ \citep[see e.g.][]{Gammie2003,McKinney06}.
KORAL also evolves the gas entropy $s$ as a passive scalar; the latter is used as a backup quantity for computing the gas energy density if the normal inversion procedure from the simulation conserved variables fails.  In the simulations reported here we set the gas adiabatic index to $\Gamma=13/9$, which lies in between the usual $5/3$ for a monatomic non-relativistic gas and $4/3$ for an ultra-relativistic gas. For this choice of $\Gamma$, the gas pressure is $p=(\Gamma-1)u = (4/9)u$. In Appendix \ref{sec:adiab}, we compare results for our $a_*=0$ model with the fiducial $\Gamma=13/9$ to simulations with the same grid and initial conditions but with adiabatic indices set to $\Gamma=4/3$ and $\Gamma=5/3$. The adiabatic index does not seem to influence the results.

The simulations here evolve only the equations of ideal GRMHD and neglect effects such as plasma resistivity \citep[e.g.][]{Ripperda2019}, radiative cooling and feedback \citep[e.g.][]{Sadowski2014,McKinney2014,Sadowski_3D,Ryan18,Morales2018,ChaelM87,Yao21}, and two-temperature evolution of separate electron and ion populations \citep[e.g.][]{Ressler15,Sadowski_2016,Dexter20}.
In particular, radiative cooling can become important in determining temperatures and potentially disk structures for hot accretion flows with accretion rates $\dot{M} \gtrsim 10^{-6}\dot{M}_{\rm Edd}$, 
(where $\dot{M}_{\rm Edd}$ is the Eddington rate), 
as in the case of the supermassive black hole in M87 \citep{EHTVIII}.

We frequently make use of the plasma-$\beta$ parameter, $\beta = 2p/b^2$ (in dimensionless code units), to characterize the ratio of the thermal pressure to the magnetic pressure,  and the magnetization parameter, $\sigma_{\rm M} = b^2/\rho$, to characterize the ratio of the magnetic energy density to the rest mass energy density. Throughout, we use the gravitational time scale $t_{\rm g} = GM/c^3$ and distance scale $r_{\rm g}=GM/c^2$ to scale quantities. Both are equal to the BH mass $M$ in natural units with $G=c=1.$

\subsection{Simulation Setup}

We have run simulations for nine different values of the BH spin parameter: $a_*=0.9$, 0.7, 0.5, 0.3, 0, $-0.3$, $-0.5$, $-0.7$, $-0.9$. The coordinate grid is modified from standard Kerr-Schild coordinates so as to concentrate resolution in both the jet region close to the polar axis and the disk region near the equatorial plane. To achieve this, we use the coordinate transformation from spatial simulation ``code coordinates" $(x_1,x_2,x_3)$ to Kerr-Schild grid coordinates $(r,\theta,\phi)$ introduced in \citet{Ressler17}.  The simulation grid grows exponentially in radius $r$ and is uniform in azimuthal angle $\phi$, while the polar angle $\theta$ is a complicated function of both $x_1$ and $x_2$, designed to concentrate resolution in the jet and disk regions. Each simulation has a resolution of $288\times192\times144$ cells in the $r$, $\theta$, and $\phi$ directions, respectively. The inner radial boundary $r_{\rm min}$ was chosen to ensure that there were 6 radial cells inside the BH horizon, and the outer boundary was fixed at $10^5\,r_{\rm g}.$

We set the following parameters for the azimuthal grid, where we use the same notation as in \citet[][Appendix B]{Ressler17}. The hyperexponential break radius is $r_{\rm br}=5000r_{\rm g}$, the collimation radii are $r_{\rm coll,jet}=1000r_{\rm g}$,  $r_{\rm coll,disk}=20r_{\rm +}$, the decollimation radii are $r_{\rm decoll,jet}=r_{\rm decoll,disk}=2r_{\rm +}$. The power-law indices are $\alpha_{\rm 1}=1$, $\alpha_{\rm 2}=0.25$. The fraction of the angular resolution concentrated in the jet and disk are $f_{\rm jet}=0.3$, $f_{\rm disk}=0.4$. We 'cylindrify' angular cells close to the axis at small radius by expanding their size in $\theta$ \citep{Tchekhovskoy2011,Ressler17}; the cylindrification radius $r_{\rm cyl}=30r_{\rm g}$ and $n_{\rm cyl}=1$. The polar angle code coordinate $x_2$ extends from
$x_{\rm 2,min}=10^{-5}$ to $x_{\rm 2,max}=1-10^{-5}$, where $x_2=0$ and 1 correspond to the two polar axes.

In Fig.~\ref{fig:grid}, we show a poloidal slice of the time- and azimuth-averaged gas density in the spin $a_*=0$ simulation, along with gridlines indicating the shape of the simulation grid in the poloidal plane.

\begin{table*}
\centering
\renewcommand{\arraystretch}{1.3}
\begin{tabular}{c | c c c c | c c c}
\hline\hline
\vspace*{0mm}

Model & $r_{\rm in}$ & $r_{\rm max}$ & $\rho_{\rm max}$& $\Gamma$ & $r_{\rm mag}$ & $A_{\phi,\rm cut}$& $\beta^{-1}_{\rm max}$\\
BH spin $a_*$ &$(r_{\rm g})$&$(r_{\rm g})$&(Arb. units)&&$(r_{\rm g})$&(Arb. units)&\\
\hline
0.9 & 20 & 41.96 & 1 & 13/9 & 400 & 0.2 &100\\
0.7 & 20 & 42.05 & 1 & 13/9 & 400 & 0.2 &100\\
0.5 & 20 & 42.15 & 1 & 13/9 & 400 & 0.2 &100\\
0.3 & 20 & 42.25 & 1 & 13/9 & 400 & 0.2 &100\\
0. & 20 & 42.43 & 1 & 13/9 & 400 & 0.2 &100\\
-0.3 & 20 & 42.62 & 1 & 13/9 & 400 & 0.2 &100\\
-0.5 & 20 & 42.75 & 1 & 13/9 & 400 & 0.2 &100\\
-0.7 & 20 & 42.9 & 1 & 13/9 & 400 & 0.2 &100\\
-0.9 & 20 & 43.06 & 1 & 13/9 & 400 & 0.2 &100\\
\hline

\end{tabular}
\caption{Parameters determining the initial \citet{Fishbone1976} tori and the poloidal magnetic field loop set by the vector potential $A_{\phi}$ in \autoref{eq:aphi}.} 
\label{tab:init}
\end{table*}

All nine simulations described here were run for long durations, $t\gtrsim 100,000\,t_{\rm g}$. This is nearly an order of magnitude longer than many other simulations reported in the literature (e.g. \citealt{Porth2019}, though there are a few that are significantly longer \citep[e.g.,][]{Narayan2012,Sadowski2013a,White2020}. 
Long-duration simulations require the initial gas supply to be sufficiently large such that there is enough gas for vigorous accretion on to the BH even at late times. 

We initialized the present simulations with spatially extended \citet{Fishbone1976} geometrically thick equilibrium tori. This torus solution is fixed by four parameters: the inner edge of the torus in the equatorial plane $r_{\rm in}$, the location of the pressure maximum (also in the equatorial plane) $r_{\rm max}$, the adiabatic index of the fluid $\Gamma$, and the maximum density $\rho_{\rm max}$. The location of the torus outer edge is sensitive to the choice of $a_*$, $r_{\rm in}$, $r_{\rm max}$, and $\Gamma$. In the present simulations, we set the inner edge of the initial torus for all the simulations at $r_{\rm in} = 20r_g$ and adjusted the radius of the pressure maximum $r_{\rm max}$ in the range $\sim42-43\,r_g$ (depending on $a_*$) such that the outer edge of the torus was at $r_{\rm out} \approx 10^4r_g$. 
We set $\rho_{\rm max}=1$, as the density normalization in GRMHD is arbitrary. The parameters of the initial tori are listed in \autoref{tab:init}.

To build up magnetic field around the BH to its saturation value, such that the accretion flow becomes magnetically arrested, we initialized the torus with a single large magnetic field loop centered around $r\approx350\,r_{\rm g}$. The loop is defined by the poloidal vector potential $A_{\phi}$:
\begin{align}
    A_{\phi} &= \mathrm{Max}\left[0,q\right], \nonumber \\
    q&= \left[\left(\frac{\rho}{\rho_{\rm max}}\right)\left(\frac{r}{r_{\rm in}}\right)^3\,e^{-r/r_{\rm mag}}\sin^3\theta\right] -  A_{\phi,{\rm cut}}.
\label{eq:aphi}
\end{align}
For all the simulations in this paper, we set $r_{\rm mag}=400\,r_{\rm g}$ and $A_{\phi,{\rm cut}}=0.2$. After determining the field components via the vector potential in Eq.~\ref{eq:aphi},
we normalized the initial magnetic field strength in the disk, following \citet{Porth2019}, such that the maximum gas pressure and maximum magnetic pressure in the torus (which do not necessarily occur at the same location) satisfy $\beta_{\rm max} \equiv (p_{\rm gas})_{\rm max} / (p_{\rm mag})_{\rm max} = 100$. 

KORAL solves the equations of GRMHD \citep[e.g.][]{Gammie2003} by advecting conserved quantities across cell walls using a finite volume method and applying geometrical source terms that encode the effects of the BH's metric at cell centers. The fluxes at the cell walls are computed using the second-order piecewise parabolic method \citep[PPM;][]{Colella1984}.
We use outflowing boundary conditions at the inner and outer radial boundaries, and reflecting boundary conditions at the polar axes. With the latter, fluid flow across the poles can sometimes create numerical instability; we control this by replacing $u^\theta$ in the innermost two cells closest to the polar axis with a value interpolated between the value in the third cell and zero. 

GRMHD simulations tend to fail in highly magnetized regions, where the gas internal energy $u$ is very small compared to other terms in the energy-momentum tensor, especially $b^2$ (see equation \ref{eq:Tmunu}). While the code conserves total energy-momentum to machine precision, in these regions small numerical errors can push the gas part of the energy-momentum tensor into an unphysical configuration, causing the simulation to fail when the code attempts to invert the energy momentum tensor to find the fluid velocity and gas density. 
To ensure numerical stability, whenever the gas becomes too highly magnetized in any region, we artificially inject gas density there in the zero angular momentum observer (ZAMO) frame\footnote{\citet{Ressler17} note that adding gas in the ZAMO frame can introduce an artificial drag which might affect the power in the jet. However, the effect is quite small since the density floor is activated only in regions where the density and internal energy are very low, and these regions are often near the stagnation point where the gas hardly moves.} \citep{McKinney2012} so as to bring the magnetization back to a ceiling value $\sigma_{\rm M} = 100$. 

\subsection{Accretion Rate and Horizon Magnetic Flux}
\begin{figure*}
   	\includegraphics[width=.9\textwidth]{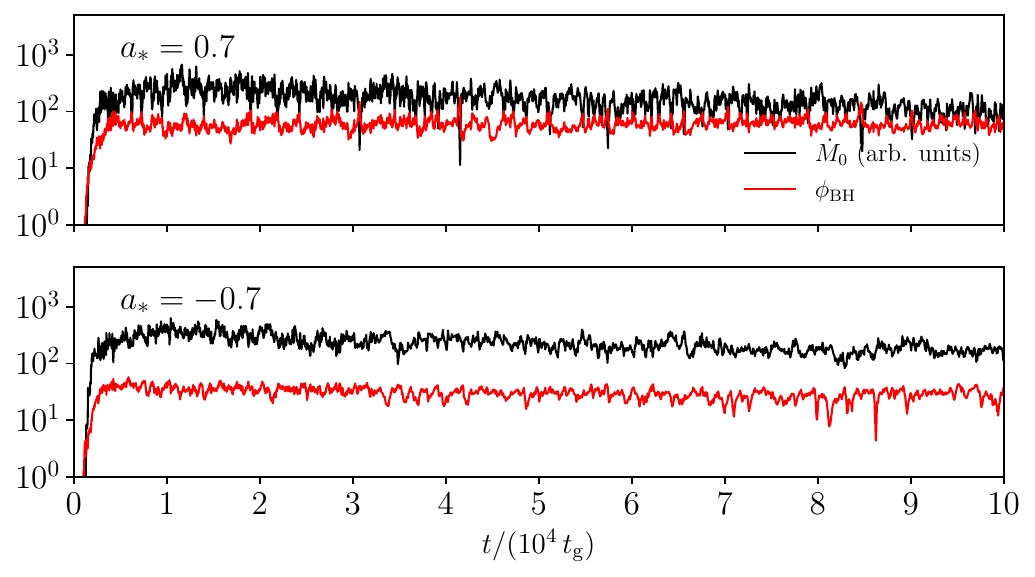}
    \caption{Top: Rest mass accretion rate $\dot{M}_0$ into the black hole (equation~\ref{eq:Mdot1}, black curve) and dimensionless magnetic flux parameter $\phi_{\rm BH}$ at the black hole horizon (equation~\ref{eq::madparam}, red) as a function of time for a simulation with $a_*=0.7$. The absolute units of $\dot{M}_0$ in GRMHD simulations are arbitrary; here, they are scaled up by a factor of 8 from the units used in the simulation to visually distinguish the $\dot{M}_0$ and $\phi_{\rm BH}$ curves. Bottom: Corresponding results for a simulation with $a_*=-0.7$. For the $a_*=0.7$ simulation, the output was saved with a cadence of $10M$ until $t=40,000M$ and with a cadence of $50M$ beyond this time. For the $a_*=-0.7$ simulation, the break in the cadence is at $30,000M$.}
    
    \label{fig:mdot_phi}
\end{figure*}

A MAD system intrinsically has a large ratio of the magnetic flux $\Phi$ through the horizon to the square root of the rest mass accretion rate $\dot{M}_0$ \citep[][]{Tchekhovskoy2011}. We compute the accretion rate as the integral of $\rho u^r$ over the horizon at $r=r_{\rm H}$:
\begin{equation}
    \dot{M}_0(t) = -\int_\theta \int_\phi \left[\rho u^r\right]_{r=r_{\rm H}} \;\sqrt{-g}\;\mathrm{d}\theta\; \mathrm{d}\phi, \label{eq:Mdot1} \\
\end{equation}
where $g$ is the metric determinant, and the negative sign is to ensure that $\dot{M}_0$ is positive when mass flows into the BH. Given $\dot{M}_0$, the dimensionless magnetic flux parameter  $\phi_{\rm BH}$, is defined to be  \citep{Tchekhovskoy2011}\footnote{We use the instantaneous accretion rate $\dot{M}_0(t)$ when computing $\phi_{\rm BH}(t)$, but we note that others \citep[e.g.,][]{Tchekhovskoy2011,McKinney2012} prefer to use a running time-averaged value of $\dot{M}_0$. In our experience, the results are similar.}
\begin{equation}
    \phi_{\rm BH}(t) = \frac{\sqrt{4\pi}}{2\sqrt{\dot{M}_0(t)}} \int_\theta \int_\phi \left|B^r\right|_{r=r_{\rm H}} \;\sqrt{-g}\;\mathrm{d}\theta\; \mathrm{d}\phi.
    \label{eq::madparam}
\end{equation}
Note that we have explicitly included a factor of $\sqrt{4\pi}$ to translate our magnetic field strength $B^r$ from Heaviside-Lorentz units to Gaussian units. Under this definition, the saturation value of $\phi_{\rm BH}$ that marks the MAD state is typically $\phi_{\rm BH}\approx50$. 

Figure \ref{fig:mdot_phi} shows the time evolution of $\dot{M}_0$ (equation~\ref{eq:Mdot1}) for two representative simulations: a prograde simulation with $a_*=0.7$, and a retrograde simuation with $a_*=-0.7$. In both simulations, by $t=10,000\,t_{\rm g}$, the accretion rate (shown by the black curves) has reached a maximum. Following the peak, there is a slow secular decline in $\dot{M}_0$ until the end of the simulation. This decline is the result of both mass-depletion of the gas reservoir and radial expansion of the initial torus from angular momentum redistribution. The decline is, however, not very extreme --- both simulations are still accreting and producing powerful jets at $t=100,000\,t_{\rm g}$. Furthermore, we normalize all our primary physical quantities by the instantaneous $\dot{M}_0$, so any slow variation of $\dot{M}_0$ with time has no effect on the results.  

Figure \ref{fig:mdot_phi} also shows the time evolution of the magnetic flux parameter $\phi_{\rm BH}$.
As defined in equation~\ref{eq::madparam}, $\phi_{\rm BH}$ is dimensionless and measures the strength of the magnetic field relative to the mass accretion rate. In both the $a_*=+0.7$ simulation and $a_*=-0.7$ simulation, $\phi_{\rm BH}$ (shown by the red curves) reaches a saturation value $\phi_{\rm BH} \sim 50$ by around $t\sim 10000\,t_{\rm g}$. Notably, $\phi_{\rm BH}$ saturates at a larger value ($\sim60$) for the prograde simulation than for the retrograde simulation ($\sim30$). This is a general trend across our sequence of simulations (noted first in \citealt{Tchekhovskoy2012}), and we comment more on it below. 

\subsection{Conserved Flux Radial profiles }

\begin{figure*}
	\includegraphics[width=8.5cm]{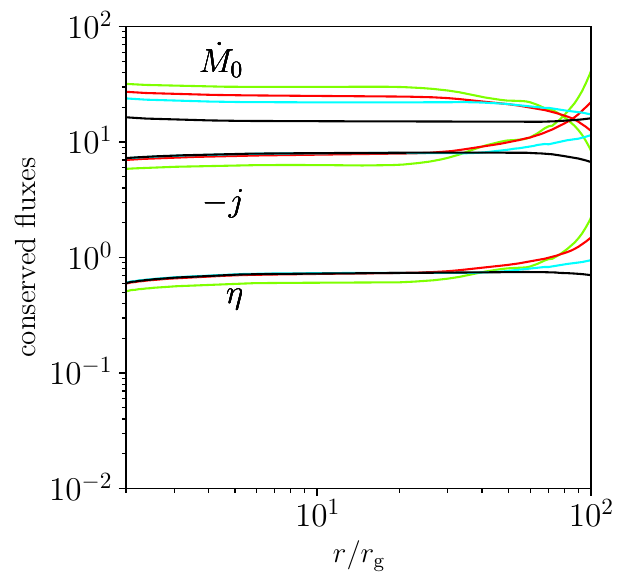}
   	\includegraphics[width=8.5cm]{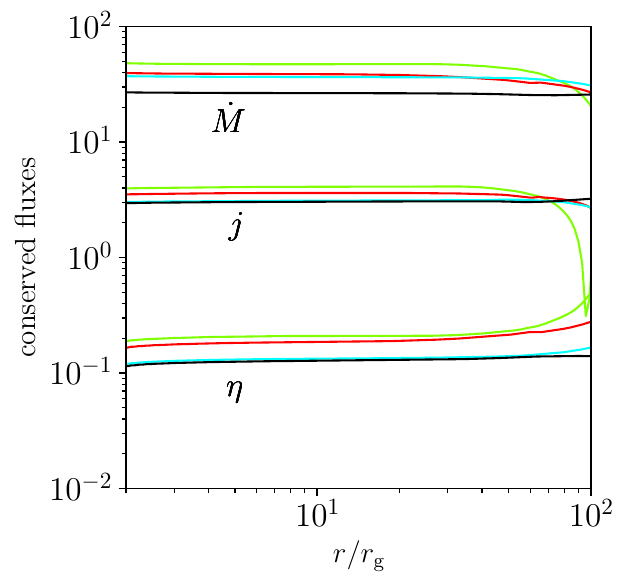}
    \caption{Left: Radial profiles of the rest mass flux $\dot{M}_0(r)$ (arbitrary units) into the black hole, the energy outflow efficiency $\eta(r)$, and the angular momentum flux into the black hole $j(r)$, for a simulation with $a_*=0.7$, averaged over four time windows:  $15,000-20,000\,t_{\rm g}$ (green curves), $20,000-30,000\,t_{\rm g}$ (red), $30,000-50,000\,t_{\rm g}$ (cyan), $50,000-100,000\,t_{\rm g}$ (black). In regions where the accretion flow has reached inflow equilibrium, the three fluxes are nearly independent of $r$. The final time window (black curves) has achieved inflow equilibrium out to almost $r\approx100r_g$. Note that the angular momentum flux into the black hole is negative in this model, i.e., the black hole loses angular momentum, while the energy outflow efficiency is quite large, $\eta\sim0.71$.  Right: Corresponding results for $a_*=-0.7$. Here the angular momentum flux into the black hole is positive, and the outflow efficiency is smaller, $\eta\sim0.13$.}
    \label{fig:fluxes}
\end{figure*}

To characterize the inward flow of energy and angular momentum in the simulations as a function of radius and time, we define the energy flux $\dot{E}$ and angular momentum flux $\dot{J}$:
\begin{align}
\dot{E}(r,t) &= \int_\theta \int_\phi T^r_{\; t}\; \sqrt{-g}\;\mathrm{d}\theta\; \mathrm{d}\phi, \label{eq:e} \\
\dot{J}(r,t) &= -\int_\theta \int_\phi T^r_{\; \phi}\;\sqrt{-g}\;\mathrm{d}\theta\; \mathrm{d}\phi. \label{eq:j}
\end{align}
The signs have been chosen such that, in each equation, the quantity measures the corresponding flux {\it into} the BH. Because the units of the density $\rho$ are arbitrary in our pure GRMHD simulations, we define a specific energy flux $e(r,t)$ and specific angular momentum flux $j(r,t)$, each normalized by the rest mass energy flux $\dot{M}_0(r,t)$:
\begin{equation}
    e(r,t) = \frac{\dot{E}(r,t)}{\dot{M}_0(r,t)} \;,\;\;\;\;\;\; j(r,t) = \frac{\dot{J}(r,t)}{\dot{M}_0(r,t)}.
\end{equation}

Note that in our definitions of the energy flux $\dot{E}$ and specific energy flux $e$, we include the flux of rest mass energy (i.e. we do not subtract out $\rho u^r$ from $T^r_{\; t}$ in the above definitions). However, in analyzing the energy flow in the simulations, we are most interested in the ratio of the `output' energy that flows out to infinity via a jet or wind to the `input' flow of rest-mass energy.  A physically useful dimensionless quantity to assess this factor is
\begin{equation}
\eta(r,t) = 1-e =\frac{P_{\rm out}}{\dot{M}_0 c^2},
\end{equation}
which measures the power $P_{\rm out}$ that escapes to infinity normalized by the rate of accretion of rest mass energy $\dot{M}_0c^2$. In principle, one should make a distinction between the total outflowing power $P_{\rm out}$ and the power in a relativistic jet $P_{\rm jet}$. However, for the MAD systems that we are studying in this paper, nearly all the energy goes into the jet, and only a small fraction of the outflowing energy goes into a non-relativistic wind, as shown by the work of \citet[][see their Figs.~9-11]{Sadowski2013a}\footnote{\citet{Sadowski2013a} defined the jet-wind boundary by the condition that the Bernoulli parameter $\mu=0.05$, which corresponds to an asymptotic outflow velocity at infinity of $0.3c$. With this definition, their jet powers were an order of magnitude (or more) larger than the wind powers. \citet{McKinney2012} defined the jet-wind boundary by the condition $b^2/\rho=1$, and found somewhat larger fractional wind power. However, even with their definition, the jet power was always significantly larger than the wind power, except when $a_*=0$ (there is no jet in this case). }. The only exception is the spin $a_*=0$ model, which has no jet. In what follows, we will refer to $P_{\rm out}$ as the jet power and the quantity $\eta$ as the jet efficiency, without subtracting out the small contribution from non-relativistic outflows. 

In Fig.~\ref{fig:fluxes}, we show radial profiles of the accretion rate $\dot{M}_0(r)$, the specific angular momentum flux $j(r)$, and the energy outflow efficiency $\eta(r)$, averaged in time over four time windows: from $15,000-20,000\,t_{\rm g}$, from $20,000-30,000t_{\rm g}$, from $30,000-50,000t_{\rm g}$, and from $50,000-100,000t_{\rm g}$. In each of these time windows, all three fluxes are constant in radius out to some radius $r_{\rm eq}$; the region $r<r_{\rm eq}$ where the fluxes are constant is considered the region of inflow equilibrium for the given time window. As expected, $r_{\rm eq}$ increases with time and reaches  its largest value, $r_{\rm eq}\approx 100\,r_{\rm g}$, in the last time window, which ends at $100,000\,t_{\rm g}$. An advantage of the long runtime of our simulations is that it gives us stable flux profiles and converged simulation properties in the disk and jet out to such relatively large radii \citep[for other long-duration simulations see, e.g.][]{Narayan2012,Sadowski2013a,White2020}.  
In the later figures in this paper, we focus primarily on the last time window from $t=50,000-100,000\,t_{\rm g}$. 

\section{Results}
\label{sec:results}
\subsection{The MAD Limit and Jet Efficiency}

\begin{figure*}
	\includegraphics[width=8cm]{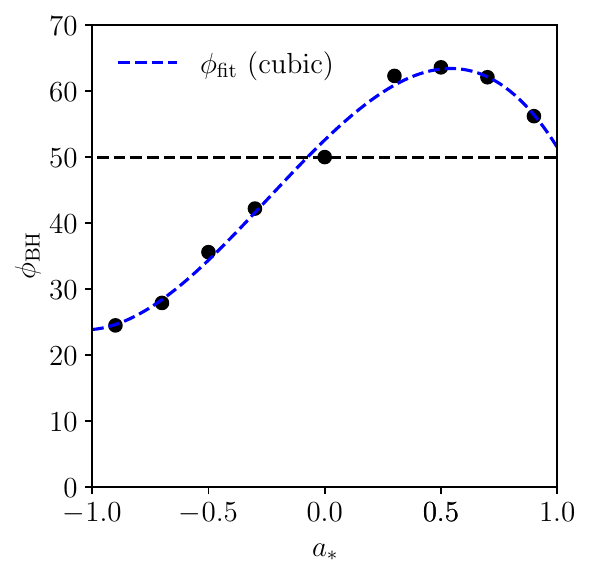}
	\includegraphics[width=8cm]{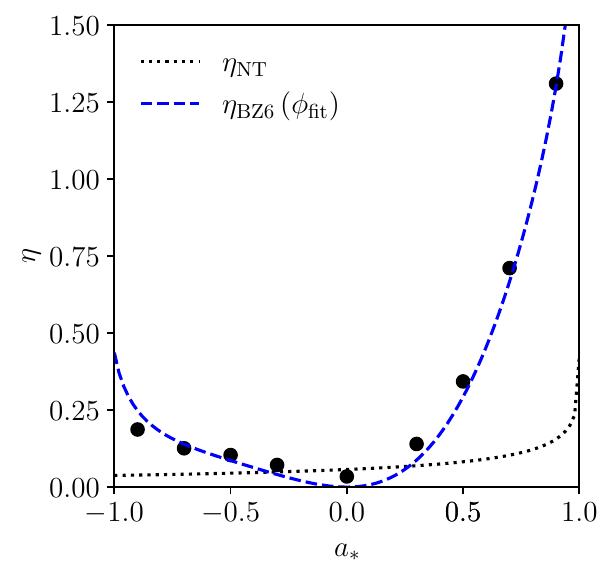}
	
    \caption{(Left) We show the time-averaged saturated magnetic flux parameter $\phi_{\rm BH}$ as a function of black hole spin $a_*$ (black dots) for the nine simulations described in this paper. The horizontal dashed line, $\phi_{\rm BH}=50$, is typically taken as the saturation value of the magnetic flux parameter, but note that $\phi_{\rm BH}$ is substantially smaller for counter-rotating disks ($a_*<0$). The dashed blue line is a third-order polynomial fit to $\phi_{\rm BH}(a_*)$ (equation~\ref{eq:phifit}).
    (Right) We show the outflow energy efficiency factor $\eta = P_{\rm out}/\dot{M}_0c^2$ (black dots). For $a_* \gtrsim 0.8$, we find $\eta>1$, which means that the jet power exceeds the entire rest mass energy flow $\dot{M}_0c^2$ into the black hole. For a given $\dot{M}_0$, the jet is much less powerful in the case of a counter-rotating disk. The dotted black line shows the efficiency of a standard \citet{NovikovThorne} thin accretion disk. The dashed blue line is the BZ6 (equation \ref{eq:BZ6}) prediction for the jet power \citep[from][]{Tchekhovskoy2010}, obtained by substituting the fitting function for magnetic flux $\phi_{\rm fit}(a_*)$ from the left panel. The result agrees with the simulations in \citet{Tchekhovskoy2012}. 
    }
    \label{fig:phi_eta_MAD}
\end{figure*}

The magnetically arrested state sets in when we have a quasi-equilibrium between the inward ram pressure of the accreting gas and the outward pressure of the confined magnetic field \citep{Narayan2003}. When this equilibrium is reached, the magnetic flux at the BH event horizon reaches a saturation value. \citet{Tchekhovskoy2011} showed
that a geometrically thick accretion disk around a rapidly rotating BH achieves a maximum value of $\phi_{\rm BH} ~\sim 50$ in the MAD limit. If $\phi_{\rm BH}$ temporarily exceeds the saturation value, magnetic flux tubes escape from the BH magnetosphere outwards into the disk, carrying away some magnetic flux, resulting in a drop in $\phi_{\rm BH}$. Such flux eruptions are behind some of the larger excursions in $\phi_{\rm BH}$ in Fig.~\ref{fig:mdot_phi}. 

The left panel of Fig. \ref{fig:phi_eta_MAD} shows the results we obtain for the mean $\phi_{\rm BH}$, time-averaged between $t=50,000-100,000t_{\rm g}$, for the 9 simulations described in this paper. For positive values of $a_*$, i.e., models in which the BH rotates in the same sense as the accretion flow, we find that
$\phi_{\rm BH}$ is roughly $\sim 60$, reducing to $\sim 50$ for $a_*=0$. Interestingly, $\phi_{\rm BH}$ continues to decline for $a_*<0$ (counter-rotating disks), falling to as low as $\sim25$ for $a_* = -0.9$. This variation of $\phi_{\rm BH}$ as a function of spin $a_*$ is very similar to the results reported in \citet{Tchekhovskoy2012}, although we use a different code and measure $\phi_{\rm BH}$ after evolving the simulation for three times longer duration. Thus, the trend of the saturation value of $\phi_{\rm BH}$ with spin shown in Fig.~\ref{fig:phi_eta_MAD} is likely a robust feature of hot MAD disks. To quantify the trend, we fit a third order polynomial (the blue dashed line in the left panel of Fig.~\ref{fig:phi_eta_MAD}): 
\begin{equation}
    \phi_{\rm fit}(a_*) = -20.2a^3_* - 14.9a^2_* + 34a_* + 52.6 \;,\; (-1\leq a_*\leq 1).
    \label{eq:phifit}
\end{equation}

A question one might ask is whether the $a_*=-0.7$ and $-0.9$ simulations, which have the lowest values of $\phi_{\rm BH}$, might have had insufficient magnetic flux in the initial torus and whether this is why $\phi_{\rm BH}$ is lower in these models. To answer this question, in Appendix \ref{sec:MAD_phibh} we present a test where we repeat the $a_*=-0.7$ simulation with a significantly stronger initial magnetic field. We find that the saturation level of $\phi_{\rm BH}$ is unaffected.

A notable feature of hot accretion flows in the MAD state is that they produce relativistic jets with power $P_{\rm out}$ comparable to, or even exceeding, the total rate of accreted rest mass energy $\dot{M}_0c^2$. 
The right panel of Fig.~\ref{fig:phi_eta_MAD} shows the time-averaged efficiency $\eta$, measured at radius\footnote{Here, and in a few other places, we choose to measure quantities at $5r_g$ rather than at the horizon. This is because GRMHD simulations can sometimes hit density floors at smaller radii, which can bias results.} $r=5r_g$, as a function of BH spin $a_*$. We find that the jet efficiency is largest for prograde disks around the most rapidly spinning BHs 
and decreases as the spin decreases. 

Figure \ref{fig:phi_eta_MAD} further reveals that, whereas high-spin ($a_*=0.9$) prograde MADs have jet efficiencies $\eta \sim 130\%$, similar high-spin retrograde MADs produce weaker jets with $\eta$ only $\sim 20\%$. It is encouraging that this behaviour is qualitatively similar to that seen in the MAD simulations of \citet{Tchekhovskoy2012}. The present study used higher resolution and the simulations were run longer, so the agreement suggests that the $\eta-a_*$ relationship displayed in Fig.~\ref{fig:phi_eta_MAD} is converged at the grid resolution used in present GRMHD simulation, which is typical for present-day simulations. Note that, for both prograde and retrograde MAD disks, the jet efficiencies are much higher than the outflow power from a standard thin accretion disk \citep{NovikovThorne}, which is shown by the dotted line in Fig.~\ref{fig:phi_eta_MAD}. 

\citet{Tchekhovskoy2010} carried out GRMHD simulations of magnetized jets confined inside a funnel-shaped rigid wall, and found that the energy efficiency was well-fitted by the following expression:
\begin{equation}
    \eta_{\rm BZ6} = \frac{\kappa}{4\pi}\phi^2_{\rm BH}\Omega^2_{\rm H}\left[1 + 1.38\Omega^2_{\rm H} - 9.2\Omega^4_{\rm H}\right],
\label{eq:BZ6}
\end{equation}
where $\Omega_{\rm H}\equiv a_*/2r_{\rm H}$ is the angular velocity of the horizon and $\kappa$ is a constant whose  precise value depends on the initial field geometry. The jet efficiency expression in equation~(\ref{eq:BZ6}) is an extended version of the traditional spin-squared dependence of the jet efficiency, $\eta_{\rm BZ}\propto a_*^2$, in \citet{Blandford1977}. 
The blue dashed line in the right panel of Fig.~\ref{fig:phi_eta_MAD} shows the prediction for the energy outflow efficiency when we substitute our fitting function for $\phi_{\rm fit}(a_*)$ (equation \ref{eq:phifit}) in the BZ6 efficiency formula (equation \ref{eq:BZ6}) with $\kappa=0.05$ (this value corresponds to the split-monopole solution). The agreement is very good.

Since radiatively inefficient accretion flows like those simulated here are found only at low mass accretion rates below about $10^{-2}-10^{-3}$ of the Eddington rate \citep{Yuan2014}, the high jet efficiencies in Fig.~\ref{fig:phi_eta_MAD} may explain why powerful jets are seen in many low-luminosity AGN and hard-state X-ray binaries. Additionally, as also noted in \citet[][]{Tchekhovskoy2010,Tchekhovskoy2012}, the steep dependence of jet efficiency on the BH spin, and also the difference between retrograde and prograde accretion disks, could explain the presence of radio-quiet and radio-loud AGNs. Note, however, that the variation of $\phi_{\rm BH}$ and $\eta$ as a function of BH spin, and the relative efficiency of prograde vs retrograde disks, as obtained from our GRMHD simulations and shown in Fig~\ref{fig:phi_eta_MAD}, are exactly opposite to the proposal in \citet{Garofalo2009}. This point has been emphasized by \citet{Tchekhovskoy2012} and \citet{Tchekhovskoy2012b}.

\subsection{Disk and Jet Structure}

\begin{figure*}
	\includegraphics[width=.6\textwidth]{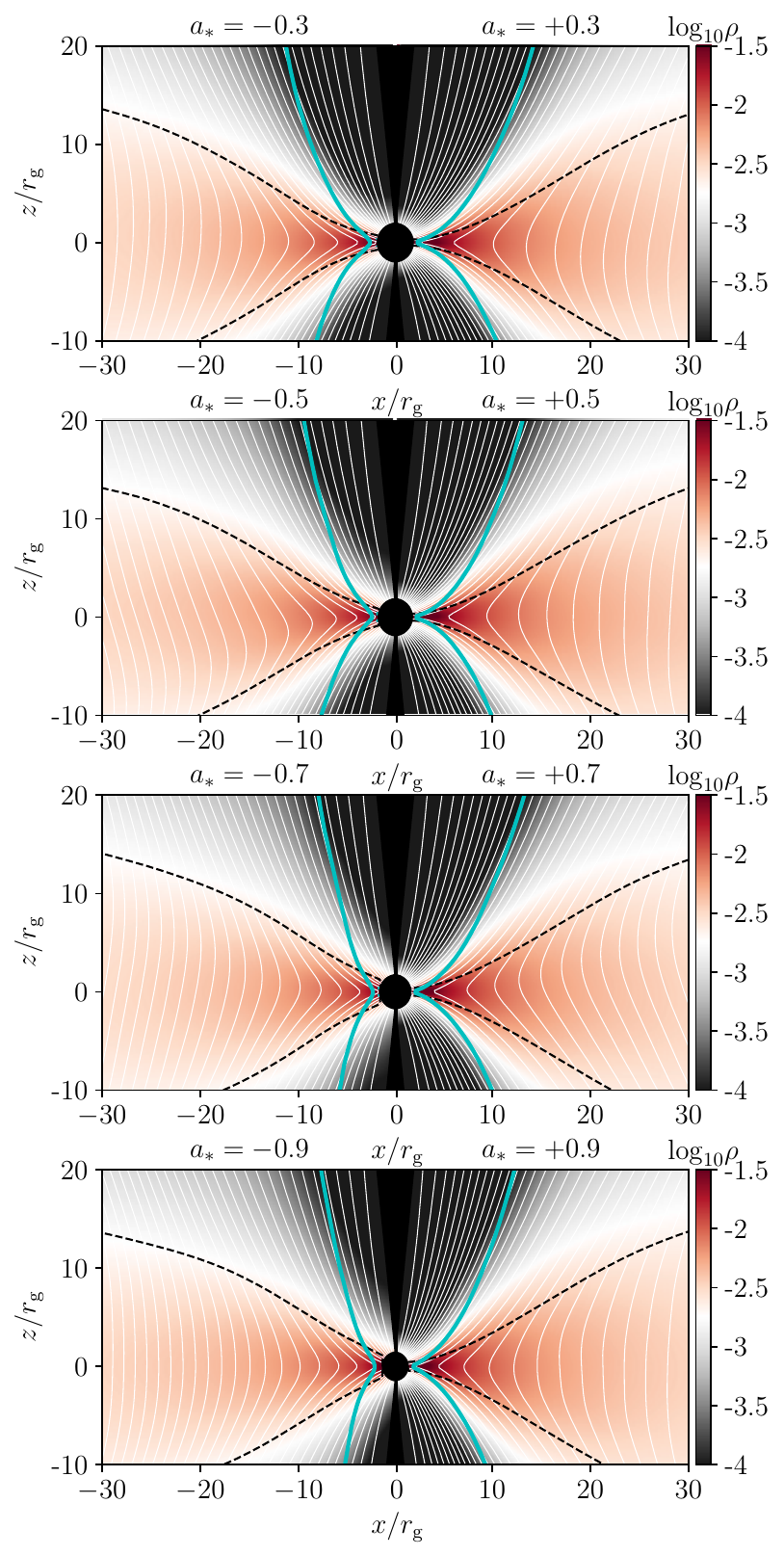}
    \caption{Time- and azimuth-averaged distributions of gas density in the poloidal plane for the eight simulations with nonzero black hole spin. The absolute value of the spin increases from 0.3 in the top row to 0.9 in the bottom row; the left side of each row shows the retrograde (negative spin) case and the right side shows the prograde (positive spin) case. In each panel, the time- and azimuth-averaged poloidal magnetic field lines are indicated with the white contours. The black dashed contour indicates the disk scale height, and the cyan contour indicates the $\sigma_{\rm M}=1$ surface, the nominal boundary of the jet. For all values of the black hole spin, the jet width in the prograde simulation is noticeably larger than in the corresponding retrograde simulation. Correspondingly, the disk scale height is smaller.}
    \label{fig:2Dprofs1}
\end{figure*}

\begin{figure*}
	\includegraphics[width=.6\textwidth]{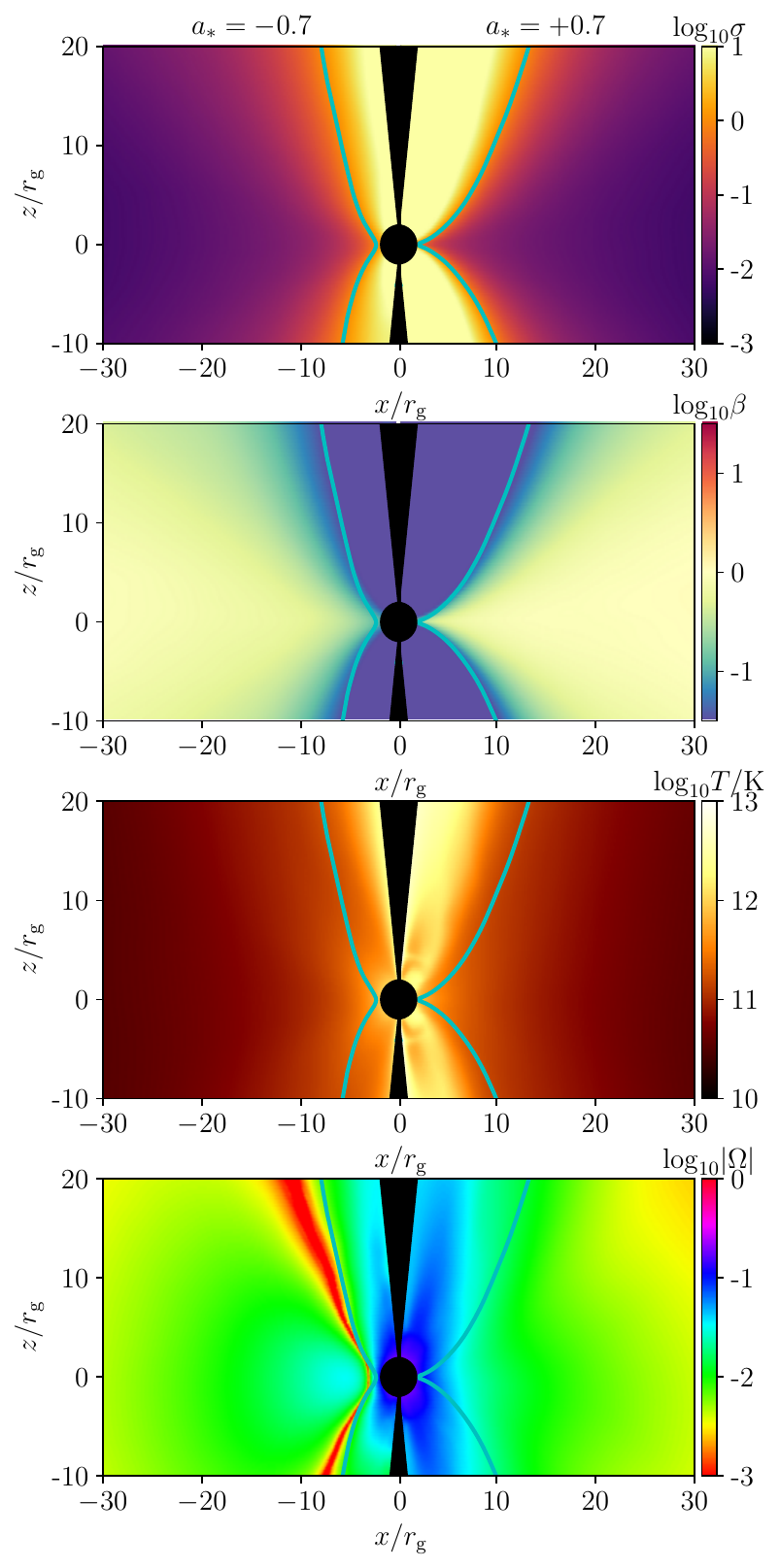}
    \caption{Time- and azimuth-averaged profiles of several quantities for the $a_*=-0.7$ retrograde simulation (left) and the $a_*=0.7$ prograde simulation (right). From top to bottom, the quantities shown are the magnetization $\sigma_{\rm M}$, the ratio of the gas to the magnetic pressure $\beta$, the gas temperature $K$ in Kelvin (assuming single-temperature fully-ionized hydrogen), and the absolute value of the angular velocity $\Omega \equiv u^\phi/u^t$. The $\sigma_{\rm M}=1$ surface is indicated in each panel by the cyan contour. In both simulations, the distributions of $\sigma_{\rm M}$ and $\beta$ transition from gas-dominated low (high) values in the disk to magnetically-dominated high (low) values in the jet at approximately the same location; this transition contour (indicated here by $\sigma_{\rm M}=1$) is further away from the pole in the prograde simulation than in the retrograde case, indicating that the jet is wider in the former case. In the retrograde simulation, the sign of $\Omega$ changes (indicated by the low values of $|\Omega|$ in the lower left panel) at approximately the same location.}
    \label{fig:2Dprofs2}
\end{figure*}

\begin{figure*}
	\includegraphics[width=5.8cm]{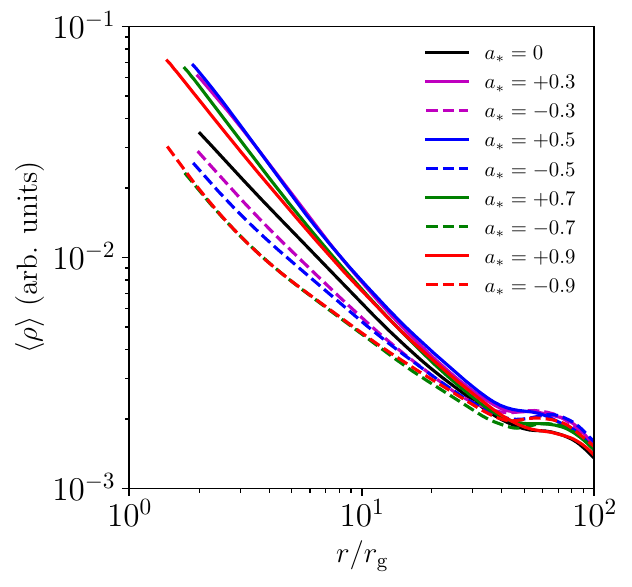}
	\includegraphics[width=5.8cm]{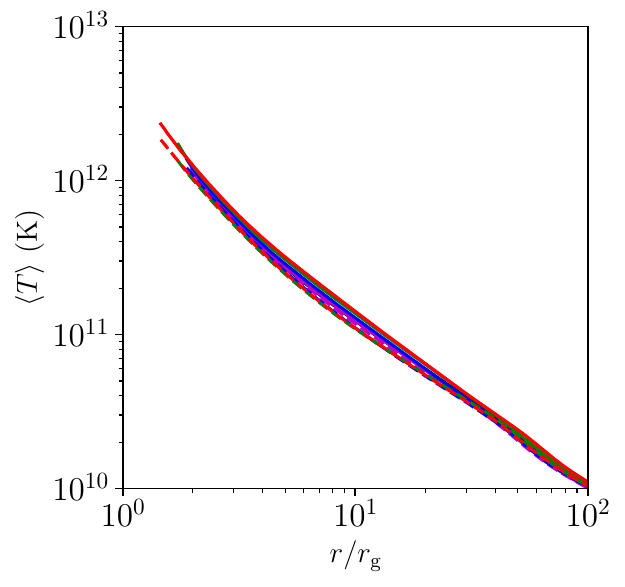}
	\includegraphics[width=5.8cm]{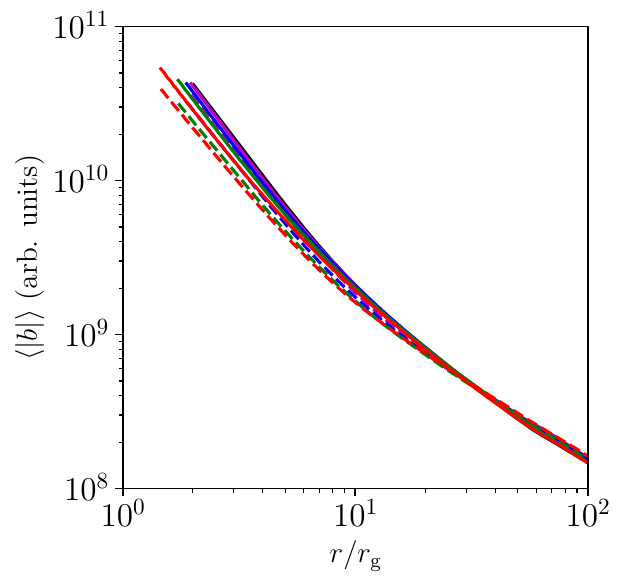}

	\includegraphics[width=5.8cm]{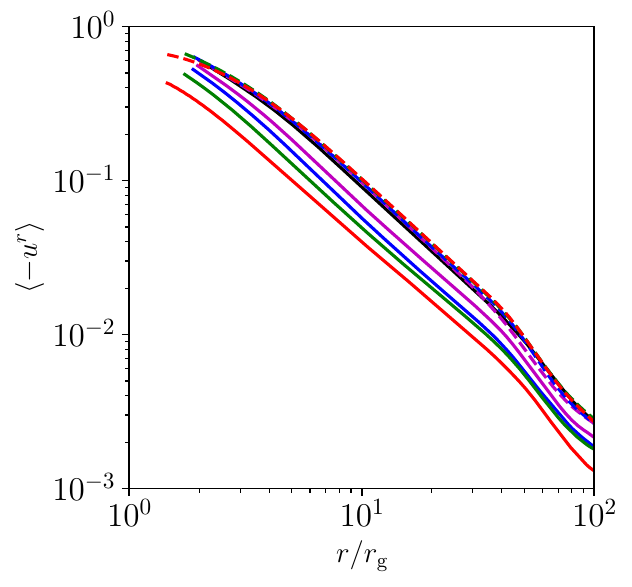}
	\includegraphics[width=5.8cm]{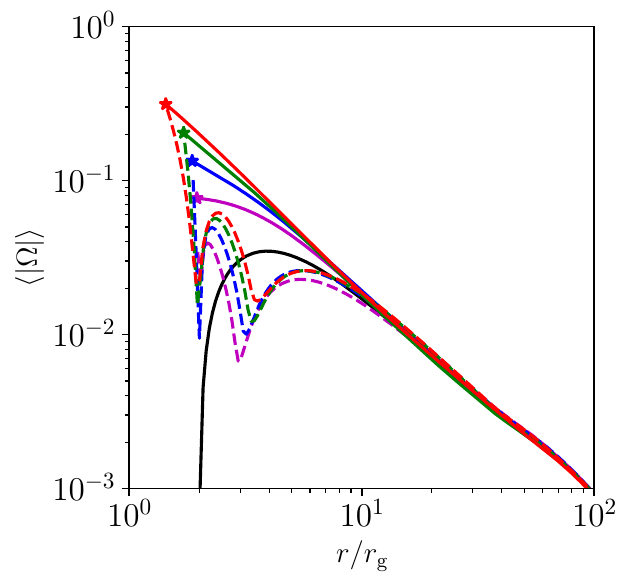}
	\includegraphics[width=5.8cm]{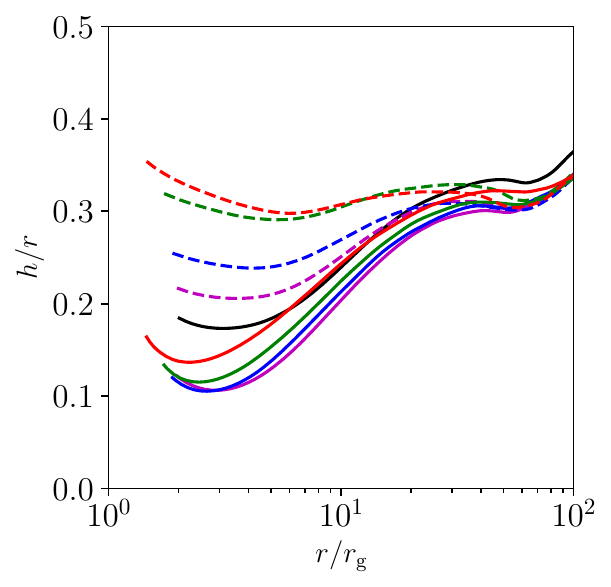}
    \caption{In the top row we show the radial profiles of the density $\langle\rho\rangle$, the temperature $\langle T \rangle$ and the magnetic field strength $\langle|b|\rangle$. 
    The averages of all quantities are density-weighted (equation~\ref{eq:otheravg}) and time-averaged between 50000 and 100000 $t_{\rm g}.$ In the bottom row we show
    the radial inward velocity $\langle-u^r\rangle$,  the angular velocity $\langle|\Omega|\rangle$, and the disk scale height ratio $h/r$ (equation~\ref{eq:hr}). 
    Line types and colors are as follows: $a_*=0.9$ (solid red curve), 0.7 (solid green), 0.5 (solid blue), 0.3 (solid magenta), 0 (solid black), $-0.3$ (dashed magenta), $-0.5$ (dashed blue), $-0.7$ (dashed green), $-0.9$ (dashed red). 
    In the plot of $\langle|\Omega|\rangle$, we indicate the angular velocity of the horizon $\Omega_{\rm H}\equiv a_*/2r_{\rm H}$ for each simulation by the star marker. }
    \label{fig:radial}
\end{figure*}

In Figs.~\ref{fig:2Dprofs1} and \ref{fig:2Dprofs2}, we show distributions in the poloidal plane of several quantities of interest, each averaged in time between $t=50,000-100,000t_g$ and over azimuth $\phi$ (from 0 to $2\pi$).  Figure~\ref{fig:2Dprofs1} shows the rest mass density $\rho$ and poloidal magnetic field lines for the eight simulations with nonzero spin. For each simulation, we indicate with a cyan line the $\sigma_{\rm M}=1$ contour (computed using the time-averaged $\rho$ and time-averaged $|B|^2$), where the magnetic field energy density equals the rest mass energy density. We take this contour as a proxy for the boundary of the magnetically dominated, relativistic jet. We also show the disk scale height $h$ (dashed black contours) as a function of radius $r$. We follow \citet[e.g.][]{Porth2019} and define the disk scale height ratio $h/r$ as: 
\begin{equation}
    \frac{h}{r} = \frac{\int\int\int \rho|\pi/2-\theta| \sqrt{-g}\,\mathrm{d}\theta\,\mathrm{d}\phi\,\mathrm{d}t}{\int\int\int \rho \sqrt{-g}\,\mathrm{d}\theta\,\mathrm{d}\phi\,\mathrm{d}t},
    \label{eq:hr}
\end{equation}
where the time average is taken over the window $t=50,000-100,000t_g$.

From Fig.~\ref{fig:2Dprofs1}, we immediately observe that the jet, defined as the region where $\sigma_{\rm M} \geq 1$, is wider in each prograde simulation relative to the corresponding retrograde simulation. The difference is most apparent for the highest spin simulations, $a_*=\pm0.9$. Conversely, close to the BH, the disk scale height $h/r$ is smaller in the prograde simulations compared to the corresponding retrograde simulations. 

Figure \ref{fig:2Dprofs2} shows several other time- and azimuth-averaged quantities in the poloidal plane for the simulations with $a_*=-0.7$ (left) and $+0.7$ (right). From top to bottom, the quantities shown are the magnetization $\sigma_{\rm M}$, the  plasma-$\beta$, the gas temperature $T$, and the angular velocity $\Omega\equiv u^\phi/u^t$. For both $\beta$ and $T$, the jet boundary, which we define by $\sigma_{\rm M}=1$, clearly delineates the transition between an ultra-hot  magnetically dominated flow in the jet region near the pole to a cooler, less magnetized flow in the equatorial disk region. The boundary is evident for both prograde and retrograde models. The $a_*=+0.7$ model shows some low-temperature regions in the jet close to the polar axis. These should not be interpreted as physically meaningful, as temperature evolution in high-magnetization regions of GRMHD simulations is unreliable and is strongly affected by the choice of density floor \citep[e.g.][]{Ressler15,ChaelM87}. These regions are unlikely to affect the radial profile of temperature in \autoref{fig:radial}, since the average over polar angle in these plots is density-weighted.

In the case of the angular velocity $\Omega$ (bottom panel), the jet boundary is not especially significant for the prograde model, but is more so in the retrograde model, where the azimuthal velocity changes direction (indicated in Fig.~\ref{fig:2Dprofs2} by the drop in $|\Omega|$ to near-zero) very near the $\sigma_{\rm M}=1$ surface. The jet in retrograde simulations rotates in the same sense as the BH, which is opposite to the direction of the disk angular momentum at large radii. In both prograde and retrograde systems, the boundary layer between the jet and the disk/wind tends to be unstable, causing the jet to become mass-loaded in fluctuating episodic events. The effect is seen especially clearly in the retrograde simulations described in \citet{Wong2020}.

\begin{figure*}
	\includegraphics[height=8cm]{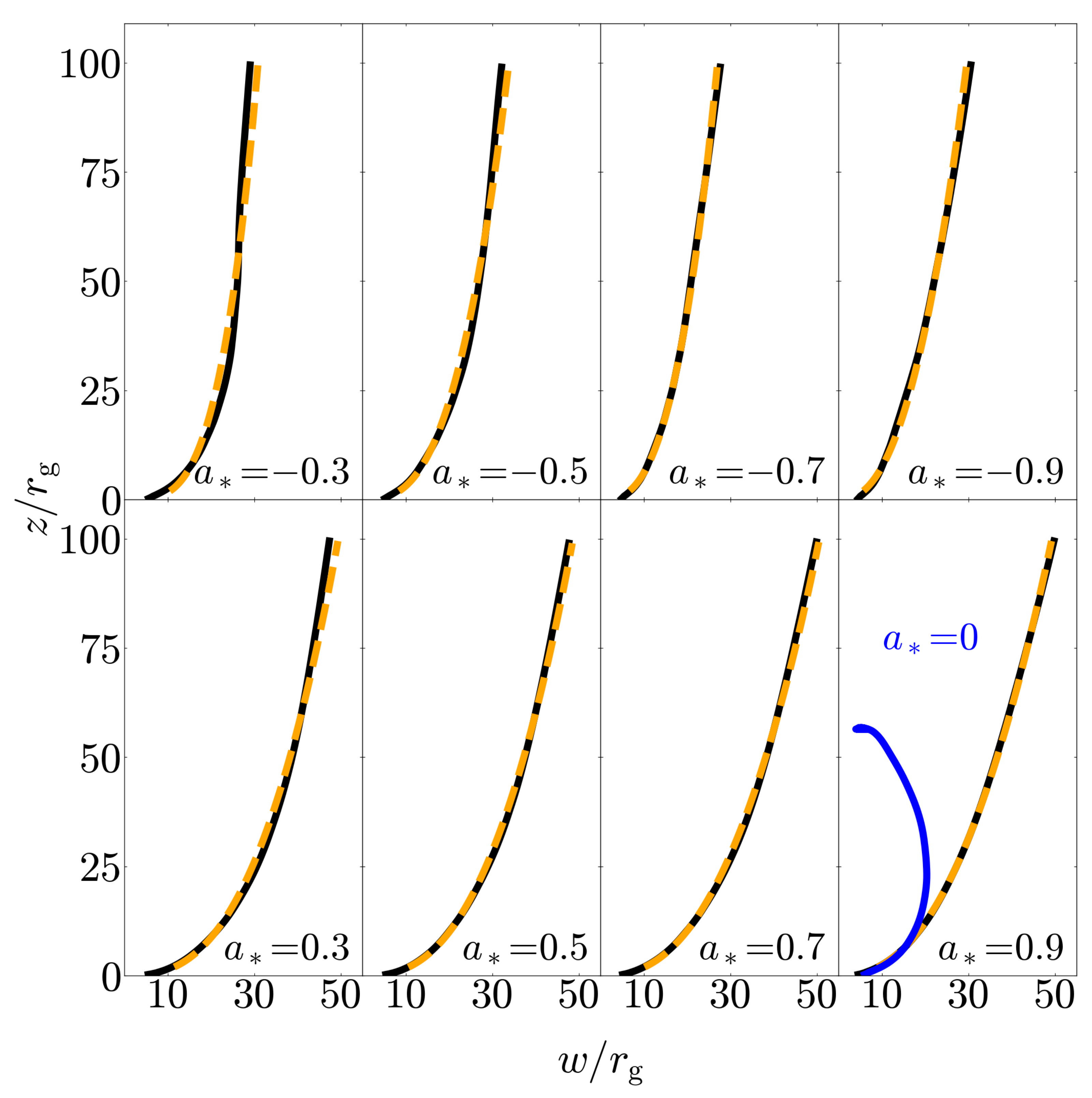}
	\includegraphics[height=8cm]{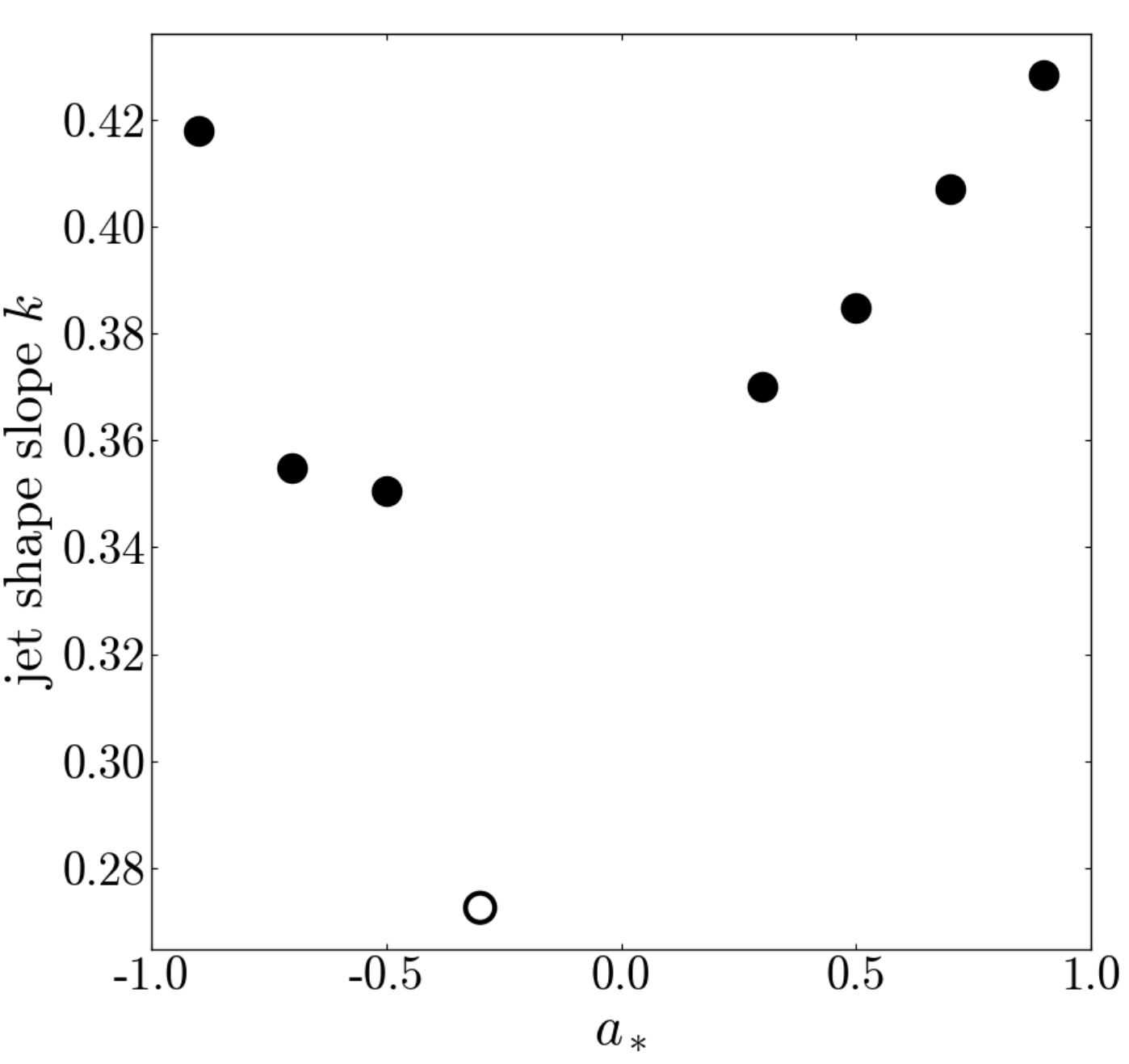}

    \caption{ (Left) Time- and azimuth-averaged jet boundary (in black), defined by magnetization $\sigma_{\rm M}=1$, for all simulations; we additionally average the jet shape over the upper and lower jet. We fit the jet shape assuming a power-law relationship between the jet width $w$ and height $z$: $w \propto z^k$ (orange-dashed). The $\sigma_{\rm M}=1$ contour for $a_*=0$ (blue, lower right panel) collapses onto the grid polar axis at $z\approx 60 \,r_{\rm g}$, indicating the lack of an extended jet. (Right) The best-fit jet shape index $k$ as a function of $a_*$. We indicate spin $a_*=-0.3$ with an open circle to indicate that the fit for $k$ is not well constrained. We do not fit for $k$ in the case of $a_*=0$.
    }

    \label{fig:jet_shape}
\end{figure*}

In Fig.~\ref{fig:radial}, we present average radial profiles of the density $\rho$, gas temperature $T$, magnetic field strength $|b| \equiv \sqrt{b^2}$, radial infall velocity $-u^r$, angular velocity $|\Omega|\equiv \left|u^\phi/u^t\right|$, and scale height ratio $h/r$ (equation~\ref{eq:hr}). 
Since we are most interested here in the behavior of these quantities in the equatorial disk, we compute the average of each quantity $q\in \left(\rho, T,|b|,-u^r, |\Omega|\right)$ weighted by density: 
\begin{equation}
     \langle q\rangle(r) = \frac{\int\int\int q\,\rho \sqrt{-g}\,\mathrm{d}\theta\,\mathrm{d}\phi\,\mathrm{d}t}{\int\int\int  \rho\sqrt{-g}\,\mathrm{d}\theta\,\mathrm{d}\phi\,\mathrm{d}t}.
    \label{eq:otheravg}
\end{equation}

The top row of Fig.~\ref{fig:radial} shows radial profiles of the density $\rho$, the temperature $T$, and the magnetic field strength $|b| = \sqrt{b^2}$. Each of the three quantities in each simulation shows a similar behaviour, an approximately broken power-law dropoff with radius, with a steeper power law slope at smaller radii $r \lesssim 5-10\,r_{\rm g}$.
The most notable difference between the simulations is seen in the density profiles for $r<10\,r_{\rm g}.$ In general, the retrograde simulations have less gas density in the innermost radii than their prograde counterparts.  However, since they have larger radial velocities and scale heights (see below), their net mass accretion rates are not very different.
(While the absolute density scale of a GRMHD simulation is not physically meaningful, each of these simulations was initialized with the same peak density in the initial torus, and deviations in the total initial torus mass are $<10$\%, almost all of it concentrated at large radii.)

The bottom row of Fig.~\ref{fig:radial} shows profiles of the radial inward velocity $-u^r$, the angular velocity $|\Omega|$, and the scale height ratio $h/r$. 
In general, prograde simulations have a smaller infall velocity $-u^r$ than retrograde models. The density-weighted angular velocity $\Omega$ switches sign in the retrograde simulations between $r \sim 3-5\,r_{\rm g}$ such that gas in the inner few gravitational radii co-rotates with the BH. 
The differences between prograde and retrograde simulations are most apparent in the scale height $h/r$. At large radii $r>10$, all simulations have similar scale heights, but at radii $r<10$ the prograde simulations all have substantially smaller $h/r\lesssim 0.1$, while the retrograde simulations have larger values $h/r \gtrsim 0.2$. This trend in $h/r$ with spin is also apparent in the poloidal profiles in Fig.~\ref{fig:2Dprofs1}. 

A notable feature of the profiles shown in Fig.~\ref{fig:radial} is the absence of any hint of the innermost stable circular orbit (ISCO). Particularly in the case of thin disks, but also to some extent in radiatively inefficient SANE accretion flows, the accreting gas has significant angular velocity and the orbital motion provides significant support against gravity. The accreting gas is thus sensitive to the loss of stable circular orbits at the ISCO, and this introduces visible features in the radial profiles of various gas properties \citep[e.g., see][for SANE disks]{Porth2019}. In contrast, the gas in the MAD systems considered in this paper is supported primarily by magnetic pressure, and rotation plays a lesser role.
As a result, there is no feature at the ISCO in any of the profiles in Fig.~\ref{fig:radial}. Of course, rotation is not totally irrelevant, since it is the sense of rotation that causes the striking differences between prograde and retrograde disks discussed earlier.

Figure \ref{fig:jet_shape} shows the time- and azimuth-averaged jet shape for all the simulations. As in Figs.~\ref{fig:2Dprofs1} and \ref{fig:2Dprofs2}, we define the jet boundary by the condition, $\sigma_{\rm M}=1$. Overall, we see that the jets in the retrograde spin models are narrower than those in the prograde models, as noted before. All jets, prograde and retrograde, exhibit generalized parabolic profiles where the width $w$ varies with vertical height $z$ as $w \propto z^k$; a dependence of this form is commonly assumed when measuring the collimation profiles of AGN jets \citep[][]{Asada2012,Kovalev2020}. In each case, the jet starts out from close to the event horizon and expands rapidly and laterally up to a few gravitational radii, beyond which the disk and the wind, with their substantially larger inertia, collimate the jet. As Fig.~\ref{fig:radial} shows,  $h/r$ is large near the BH in the retrograde simulations. A larger disk scale height results in stronger collimation, and hence a narrower jet. 

\begin{table}
\centering
\renewcommand{\arraystretch}{1.3}
\begin{tabular}{c | c c}
\hline\hline
\vspace*{0mm}

BH spin & $k_0$ & $k$\\

\hline
$0.9$  & $0.837\pm 0.001$ & $0.428\pm 0.001$\\
$0.7$  & $0.890\pm 0.001$ & $0.407\pm 0.001$\\
$0.5$  & $0.918\pm 0.001$ & $0.385\pm 0.001$\\
$0.3$  & $0.954\pm 0.002$ & $0.370\pm 0.001$\\
$-0.3$ & $0.945\pm 0.005$ & $0.273\pm 0.003$\\
$-0.5$ & $0.828\pm 0.003$ & $0.351\pm 0.002$\\
$-0.7$ & $0.723\pm 0.001$ & $0.355\pm 0.001$\\
$-0.9$ & $0.636\pm 0.003$ & $0.418\pm 0.002$\\

\hline

\end{tabular}
\caption{Fit parameters for the jet shapes in the left panel of Fig.~\ref{fig:jet_shape}, $\log_{10}w = k_0+k\,\log_{10}z$, for the spinning BH models. We fit for the jet profile between $z=5 - 100 \,r_{\rm g}$.}
\label{tab:jet_shape}
\end{table}

We calculate the collimation profile of the jet in the form, $\log_{10}w = k_0+k\,\log_{10}z$, using the Python function {\tt curve\_fit}, where the fit is limited to the range $z=5 - 100 \,r_{\rm g}$. Table~\ref{tab:jet_shape} shows the fit results for the parameters $k_0$ and $k$ for each jet model except the BH spin $a_*=0$ case.
In Fig.~\ref{fig:jet_shape}, the panel on the right shows the best-fit values for the index $k$ for all the simulations with a spinning BH. The power-law slope ranges from $k\approx 0.27 - 0.43$. These values are slightly smaller than those measured for the parsec/kiloparsec-scale jets in several AGNs, e.g., $k\approx 0.39-0.56$ for AGNs considered in \citet{Kovalev2020} and $k\approx 0.39-1.86$ from \citet{Boccardi2021}.\footnote{Note that some of the values of $k$ were measured near the transition radius from a parabolic shape to a conical or wider structure further out. Hence it is possible to have $k>1$.} In the case of M87's jet, the power-index is measured to be $k=0.57$, transitioning to $k=0.9$ at a few 10s of parsecs, which is approximately a few $\times 10^5 r_{\rm g}$ \citep[][]{Asada2012, Nokhrina2019}. We expect the value of $k$ to be slightly smaller for our simulations compared to observations since the observed radio emission in AGN jets originates in the jet sheath \citep[e.g.,][]{Kim2018, Janssen2021}, which is likely to be less collimated than the $\sigma_{\rm M}=1$ jet boundary that we consider. Modulo this caveat, we see from Fig.~\ref{fig:jet_shape} see that $k$ increases with increasing BH spin magnitude $|a_*|$ for both prograde and retrograde disks, indicating that the collimation profile depends directly on the jet power. We will discuss trends in the jet width further in Section~\ref{sec:phimad_corr}. 

A single power-law description for the collimation profile does not always work, as seen in the case of the $a_*=-0.3$ model, where the jet shape seems to require a $k$ value that varies with height. In this model, the jet is rather weak: $\eta\sim7.2\%$, which is not dissimilar to $\eta\sim3.5\%$ for the spin 0 model which has no extended jet. Parabolic jet profiles are less likely for such weak jets.

\subsection{Variability}

\begin{figure}
	\includegraphics[width=8cm]{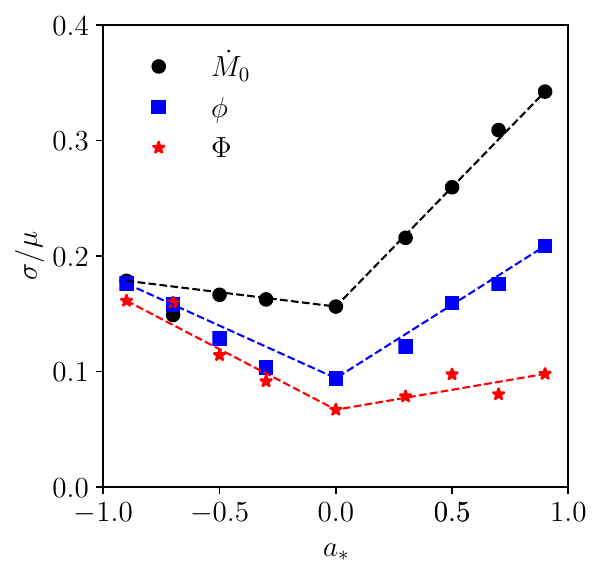}
    \caption{Variability $\sigma/\mu$ of the mass accretion rate $\dot{M}_0$  (black circles), the magnetic flux parameter $\phi_{\rm BH}$ (blue squares), and horizon magnetic flux $\Phi$ (red stars), plotted as a function of the black hole spin $a_*$ for the nine simulations. The quantities are all calculated at the black hole event horizon. The dashed lines connect the points at $a_*=0$ to the points at $a_*=0.9$ and $-0.9$, and are meant to highlight the trends.}
    \label{fig:var}
\end{figure}

Figure \ref{fig:var} shows the variability of the mass accretion rate $\dot{M}_0$, the dimensionless magnetic flux parameter $\phi_{\rm BH}$, and the horizon magnetic flux $\Phi$ (this is the integral in equation~\ref{eq::madparam} without the normalizing pre-factor), for the nine simulations. The variability is computed over the time range $50000-100000\,t_{\rm g}$. We sub-divided this time range into 50 bins of duration $1000\,t_{\rm g}$, and for each time bin and each quantity $q$, we calculated the mean $\mu$ and its variance around the mean $\sigma^2$ as follows,
\begin{equation}
\mu = \frac{1}{n} \sum_{i=1}^n q_i, \qquad \sigma^2 = \frac{1}{(n-1)} \sum_{i=1}^n (q_i-\mu)^2, 
\end{equation}
where $n$ is the number of samples in the given time bin. The ratio $\sigma/\mu$, averaged over the 50 bins, provides a dimensionless measure of the variability on time scales shorter than the bin size of $1000\,t_g$. 

The black dots in Figure \ref{fig:var} show the average ``modulation index" $\sigma/\mu$ for $\dot{M}_0$ as a function of BH spin (the values are listed in Table~\ref{tab:summary}). We see that $\sigma/\mu$ is substantially larger for the models with positive $a_*$, which also have larger values of $\phi_{\rm BH}$ (Fig. \ref{fig:phi_eta_MAD}), compared to the models with $a_* \leq 0$. Interestingly, the retrograde spin models all show similar values of $\sigma/\mu$. 

The variability in $\Phi$ (red stars in Fig.\ref{fig:var}) exhibits the opposite trend as in $\dot{M}_0$, with modulation index $\sigma/\mu$ increasing with the retrograde spin magnitude while remaining largely independent of prograde spin. This opposite behavior suggests that the transport of mass and of magnetic flux are mostly uncorrelated. Indeed, a cross-correlation analysis of the fluctuations in $\dot{M}_0$ and $\Phi$ gives a correlation coefficient of only $\sim0.2$, which is surprisingly small.

In MAD systems, flux eruptions contribute strongly to the variability, for both the accretion rate and the magnetic flux. It is possible that the strength and periodicity of flux eruptions depends strongly on the BH spin, such that for retrograde spins, eruption events are inefficient in pushing gas away but eject out magnetic flux quite readily. Additionally, the retrograde models have geometrically thicker disks close to the BH, with the disk scale height reaching $h/r\gtrsim 0.25$. Hence accretion in retrograde models may continue at higher altitudes even as bundles of vertical field lines are ejected radially outward in the disk midplane. We offer these as speculative possibilities.

Finally, we show the variability of the dimensionless magnetic flux parameter $\phi_{\rm BH}$ in Figure~\ref{fig:var} by the blue squares. In this case, $\sigma/\mu$ increases with the spin magnitude $|a_*|$ for both prograde and retrograde systems, and the variability magnitude lies in between the $\sigma/\mu$ values of $\dot{M}_0$ and $\Phi$. By cross-correlating $\phi_{\rm BH}$ with $\dot{M}_0$ and $\Phi$ we find that, for positive BH spin, fluctuations in $\phi_{\rm BH}$ are driven mostly by $\dot{M}_0$ variations. For instance, for $a_*=0.9$, the cross-correlation coefficient between $\phi_{\rm BH}$ and $\dot{M}_0$ is $-0.81$, whereas the coefficient between $\phi_{\rm BH}$ and $\Phi$ is only $0.19$. We find the opposite behavior for negative spin values. For $a_*=-0.9$
the cross-correlation coefficient between $\phi_{\rm BH}$ and $\dot{M}_0$ is $-0.41$, but the coefficient between $\phi_{\rm BH}$ and $\Phi$ is 0.82.

We have verified that the above variability results are not sensitive to our choice of a window size of $1000t_g$. We find similar results for $500t_g$ and $2000t_g$. We note that $1000t_g$ corresponds to 6 hours in the case of Sgr A* (a half night's worth of observing), and about a year for M87. These two BHs are the primary targets for the Event Horizon Telescope \citep{PaperI}. We also note that \citet{White_Chrystal2020} find the variability characteristics of simulations to be somewhat sensitive to the numerical resolution employed. The resolution in our simulations is fairly high by current standards (though not as high as the best currently in the literature, e.g., \citealt{Ripperda2021}), so we do not expect resolution to be an issue in our work.

Note that variability in $\dot{M}_0$ (or $\Phi$ or $\phi_{\rm BH}$) does not immediately translate to variability in the radiative luminosity. The latter needs to be investigated separately after including radiation physics in the simulation output and post-processing with ray-tracing software. However, we note that the 230 GHz radiation in M87* originates close to the event horizon \citep[][]{EHTV,chael2021}, and thus, the variability $\sigma/\mu$ of the 230 GHz lightcurve should follow that of $\dot{M}_0$. Indeed, \citet[][]{Chatterjee2020arXiv} showed that the variability amplitude of $\dot{M}_0$ and the ray-traced 230 GHz lightcurve for Sagittarius A* are comparable, with variability amplitude $\sigma/\mu \sim 0.24-0.31$. 

\subsection{Black Hole Spindown} \label{sec:spindown}
\begin{figure*}
	\includegraphics[width=\textwidth]{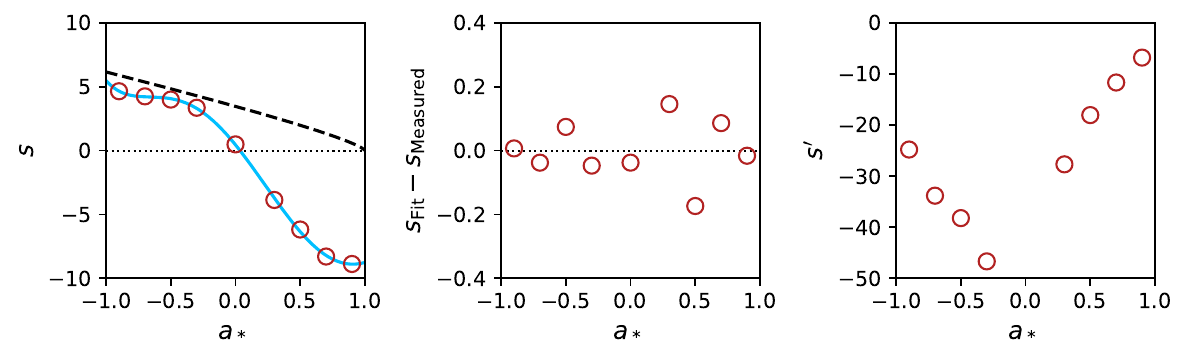}
    \caption{Left: The spinup parameter $s$ (equation \ref{eq:spinup}) as a function of the black hole spin $a_*$ for the nine simulations described in this paper.  The blue curve is a fifth degree polynomial fit to these values.  The dashed black line shows for comparison the corresponding result for a standard thin accretion disk \citep{Shapiro2005}.  Unlike a thin disk, a MAD corotating disk ($a_*>0$) causes the black hole to spin {\it down} efficiently because of angular momentum loss to the jet.  Center:  The absolute error in $s$ from the polynomial fit in the left panel. Right: The modified spinup parameter $s'$ (\autoref{eq:spinup2}), which measures the amount of spinup/spindown for a given energy output in the jet. (We omit $a_*=0$ from the final panel since this model does not have a jet.)
    \label{fig:spindown}}
\end{figure*}

The jets in the simulations described here receive their power from the spin of the BH. 
We find that in all simulations with $a_*\neq0$, more angular momentum is lost to the jet than is supplied by the accretion disk.  Consequently, the BHs lose angular momentum over time.

Following \citet{Shapiro2005}, we define the  spinup parameter $s$, 
\begin{equation}
    s \equiv \frac{da_*}{dt}\frac{M}{\dot{M}_0} = \frac{da_*}{dM_0/M} = j - 2e a_*. \label{eq:spinup}
\end{equation}
We measure $j$ and $e$ at $r=5r_g$. Standard thin accretion disk models have positive spinup parameters $s$ for all spin values up to $a_*=0.998$ \citep{Thorne1974}.  Hence, counterrotating accretion disks always spin the BH down and corotating accretion disks spin the BH up.

The left panel in \autoref{fig:spindown} shows the values of $s$ that we find for the nine simulations. All the four simulations with corotating disks around spinning BHs (the rightmost four points) have negative values of $s$. That is, in all four of these models, the BH spins {\it down} as a result of powering the jet. Note that these BHs do receive positive angular momentum from the accreting gas. However, this contribution is overwhelmed by the loss of angular momentum via the jet, and so the net effect is that the BH spins down. As in the case of our results for $\phi_{\rm BH}$ and $\eta$ as a function of $a_*$, the form of $s(a_*)$ we find is similar to that found in shorter-duration MAD simulations by \citet[][see also the ``thinner disc TNM11" models in \citealt{McKinney2012} for additional results]{Tchekhovskoy2012}.

For counter-rotating disks, $s$ is positive, i.e., the BH gains positive angular momentum (defined with respect to the accretion flow). However, since these BHs have negative angular momenta ($a_*<0$), a positive $s$ again corresponds to the spin energy of the BH decreasing with time. We thus conclude that, generically, hot accretion flows in the MAD state spin {\it down} their central BHs. 
The spin $a_*=0$ model is a special case, and shows a weak spinup as a result of the inflowing gas having non-zero angular momentum. However, the spinup in this case is far less than the equivalent rate in the case of a thin accretion disk \citep{Shapiro2005}, plotted with a dashed black line in \autoref{fig:spindown}. 

In the left panel of \autoref{fig:spindown}, we show a fifth degree polynomial fit to $s(a_*)$: 
\begin{equation}
s(a_*) \approx 0.45 - 12.53 a_* -7.80 a_*^2 + 9.44 a_*^3 + 5.71 a_*^4 - 4.03 a_*^5. \label{eq:sastar}
\end{equation}
We show the residuals of this fit in the central panel.

The parameter $s$ measures spinup normalized by the rate of accretion of rest mass energy  $\dot{M}_0$. However, $\dot{M}_0$ is generally difficult to estimate from observations. A potentially more useful way of scaling spinup is via the power $P_{\rm out}$ carried out in the jet (where $P_{\rm out} \equiv dE_{\rm out}/dt = \eta \dot{M}_0c^2$). We thus define
\begin{equation}
    s' = \frac{d|a_*|}{dE_{\rm out}/Mc^2} = \frac{s}{\eta}\,{\rm sgn}(a_*). \label{eq:spinup2}
\end{equation}
The right panel in \autoref{fig:spindown} shows how $s'$ behaves as a function of $a_*$. By this measure, for a given jet power, the spindown is fastest for a moderate retrograde spin. Note that in order to produce the same jet power, the mass accretion rate would have to be larger for low spin BHs as compared to high spin BHs. The differences between the magnitudes of $s$ and $s'$ is explained by differences in the jet efficiency (see Fig.~\ref{fig:phi_eta_MAD}).

BH spindown via accretion can have a non-negligible effect on the spin evolution of massive BHs across cosmic time; we explore the consequences in Section~\ref{sec:cosmicspindown}.

Note that equation (\ref{eq:sastar}) gives spinup-spindown equilibrium, i.e., $s=0$, at a small positive value of the BH spin $a_{\rm eq} \approx 0.035$. If a spinning BH were to accrete for an extremely long time in the MAD state, one would expect the BH spin to asymptote to $a_*=a_{\rm eq}$. However, the value of $a_{\rm eq}$ itself probably drifts with time. For instance, \citet{Tchekhovskoy2012}, whose simulations were of shorter duration, found a larger $a_{\rm eq} \approx 0.07$. We speculate that, for sufficiently long-lived MAD systems, $a_{\rm eq}\to 0$.

\section{Discussion}
\label{sec:discussion}

 \subsection{Correlations with $\phi_{\rm BH}$}
 \label{sec:phimad_corr}
 
\begin{figure*}
    \includegraphics[width=8cm]{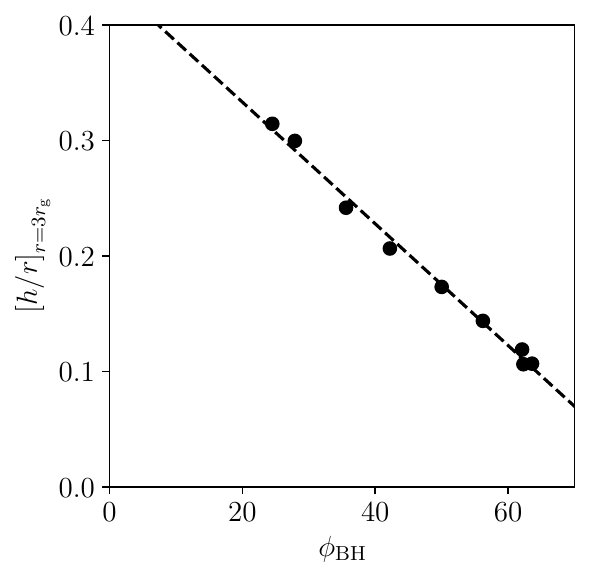}
    \includegraphics[width=8cm]{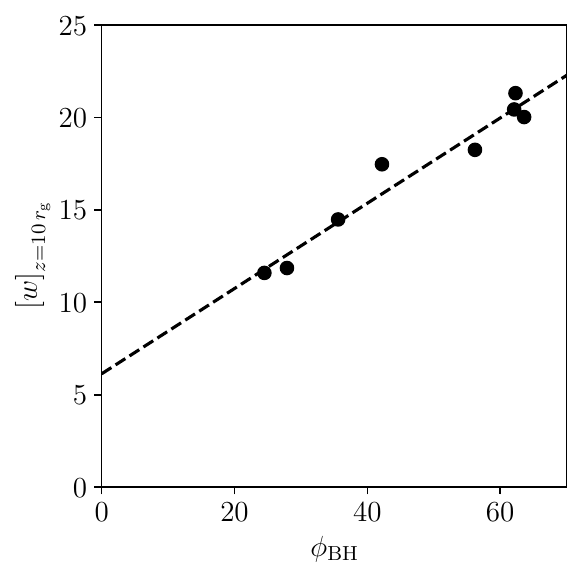}
    \caption{(Left) The scale height ratio $h/r$ vs the saturated magnetic flux parameter $\phi_{\rm BH}$ for each simulation. The disk scale height is calculated at $r=3\,r_{\rm g}$. (Right) The jet width $w_{\rm jet}$, defined as twice the cylindrical radius of the $\sigma_{\rm M}=1$ contour, from the time- and aximuth-averaged data at height $z=10 \,r_{\rm g}$. The scale height appears to decrease linearly with magnetic flux $\phi_{\rm BH}$, and the jet width to increase linearly with $\phi_{\rm BH}$. }

    \label{fig:phiMAD_profs}
\end{figure*}

In Figures~\ref{fig:2Dprofs1}, \ref{fig:radial} we showed that the disk scale height $h/r$ at small radii is smaller in prograde simulations than in the corresponding retrograde simulations; conversely, the jet width defined by the $\sigma_{\rm M}=1$ surface is larger in prograde simulations than in retrograde systems (Fig.~\ref{fig:jet_shape}). Furthermore, BHs surrounded by retrograde MADs have less magnetic flux than BHs with the same spin magnitude in prograde systems (Fig.~\ref{fig:phi_eta_MAD}). In Fig.~\ref{fig:phiMAD_profs}, we connect these observations and plot the scale height $h/r$ and jet width $w_{\rm jet}$ as a function of $\phi_{\rm BH}$. 

We compute the disk scale height using equation~\ref{eq:hr} at $r=3\,r_{\rm g}$, and define the jet width $w_{\rm jet}$ as twice the cylindrical radius  of the $\sigma_{\rm M}=1$ contour (of the $t$- and $\phi$-averaged data in  Fig.~\ref{fig:2Dprofs1}) at a height $z=10\,r_{\rm g}$.\footnote{The choices of $r$ and $z$ here are motivated by observations of the supermassive black hole in M87 by the Event Horizon Telescope \citep{EHT_I} and the GRMHD-based models that were used to interpret the observed image \citep{EHT_V}.} Note that the $a_*=0$ simulation does {\it not} have a magnetized relativistic jet; its $\sigma_{\rm M}=1$ contour does not extend to large radii but begins to close in at $r\approx30\,r_{\rm g}$ (Fig.~\ref{fig:jet_shape}). As a result, we do not compute a jet width for $a_*=0$. 

The disk scale height $h/r$ at $3\,r_{\rm g}$ decreases linearly with increasing $\phi_{\rm BH}$ (Fig.~\ref{fig:phiMAD_profs}, left panel). Conversely, the jet width $w_{\rm jet}$ at $z=10\,r_{\rm g}$ increases  nearly linearly with $\phi_{\rm BH}$. It is likely that these effects are related. A BH with more magnetic flux produces a wider magnetically dominated jet; a wider jet then compresses the equatorial disk near the BH to a thinner region around the midplane than in a system with less magnetic flux and a less powerful, narrower jet. The linear trends of $h/r$ and jet width $w_{\rm jet}$ with $\phi_{\rm BH}$ are simpler than the more complicated variations as a function of BH spin $a_*$ or jet efficiency $\eta$. This is because the relationships between $a_*$, $\eta$ and $\phi_{\rm BH}$ themselves are not linear (Fig.~\ref{fig:phi_eta_MAD}). 

The correlation of jet width and disk scale height indicated by Fig.~\ref{fig:phiMAD_profs} could have observational consequences. For instance, the jet width in M87 has been measured within $\approx30\,r_{\rm g}$ of the central BH by \citet{Hada2016} and \citet{Kim2018}. Recent EHT polarimetric results suggest that the central accretion disk in M87* is MAD \citep{EHTVIII}, so the relationship in Fig.~\ref{fig:phiMAD_profs} between jet width and $\phi_{\rm BH}$ in the present MAD simulations could potentially be used to infer the saturation magnetic flux in M87*. The inferred magnetic flux could then be used to solve for the BH spin via the relationship established in equation~(\ref{eq:phifit}). However, this measurement would face several significant sources of systematic uncertainty. First, the jet viewing angle and bulk Lorentz factor must be well-constrained to de-project the observed widths. Second, we will need to calibrate the observed jet width against the width of the $\sigma_{\rm M}=1$ surface in the simulations; in practice, the jet emission may not be brightest exactly on this contour. In addition, not all the correlations are monotonic, so there may be double-valued solutions.

Furthermore, if the accretion flow in M87* is MAD, it is likely that the 230 GHz EHT image of emission immediately surrounding the BH originates from the equatorial disk close to the horizon \citep[e.g.][]{EHTV,chael2021}. Analyses of future EHT images of the central few $r_{\rm g}$ in M87* may constrain the disk scale height $h/r$ and thus provide another handle on $\phi_{\rm BH}$ and $a_*$, assuming the relationships derived in our set of 9 simulations hold for MAD systems generally. 

\subsection{Black hole Spindown over Cosmic Time} \label{sec:cosmicspindown}

\begin{figure}
	\includegraphics[width=0.5\textwidth]{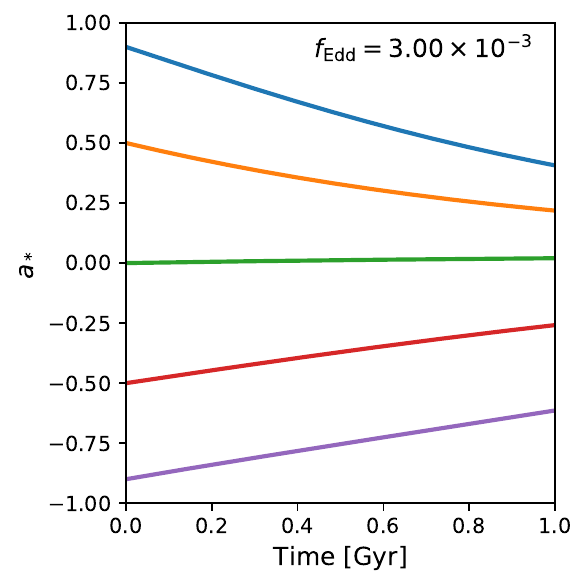}
    \caption{Spin evolution of MADs continuously accreting at an Eddington ratio of $3\times 10^{-3}$ for 1 Gyr, estimated by numerically integrating \autoref{eqn:spin_integration}.  Black hole spins are initialized at $a_* \in \{-0.9, -0.5, 0, 0.5, 0.9\}$.  While steady accretion at this level for 1 Gyr is optimistic, this exercise demonstrates that spindown in the MAD regime can be relevant for the cosmic evolution of black hole spins.
    \label{fig:spindown_evolution}}
\end{figure}

In this paper we find that the spindown of BHs from the angular momentum lost to the jet in the MAD state is significant. This may have consequences for the cosmic evolution of BHs. Using a fixed Eddington ratio and the fit for $s(a_*)$ in equation~(\ref{eq:sastar}), we numerically integrate

\begin{equation}
    \frac{da_*}{dt} = f_\mathrm{Edd} \frac{\dot{M}_\mathrm{Edd}}{M} s(a_*),
    \label{eqn:spin_integration}
\end{equation}

\noindent where the Eddington accretion rate is given by $\dot{M}_\mathrm{Edd} = (4 \pi G m_p M)/(\epsilon \sigma_{\rm T} c)$, with $\sigma_{\rm T}$ the Thomson cross-section and assuming the fiducial radiative efficiency $\epsilon$ to be 0.1. Since $\dot{M}_\mathrm{Edd} \propto M$, the mass dependence in equation~(\ref{eqn:spin_integration}) cancels out.  

In Fig.~\ref{fig:spindown_evolution}, we plot the spin evolution of BHs initialized with spins $a_* \in \{-0.9, -0.5, 0, 0.5, 0.9\}$.  We assume optimistic accretion parameters, with continuous accretion at the rate of $f_\mathrm{Edd} \equiv \dot{M}_0 / \dot{M}_{\rm Edd} =3\times10^{-3}$ (near the boundary between geometrically thin and thick disks) for 1\,Gyr.  This Eddington ratio is near the approximate boundary above which a hot accretion flow is expected to transition to a thin disk \citep{Yuan2014}.  Under the assumed conditions, we find cosmologically significant spindown, with the $a_*=0.9$ model reaching $a_*\sim 0.4$ at the end of the time period. This spindown is likely to be most relevant for BHs at the centers of massive elliptical galaxies, which are thought to steadily accrete at low Eddington rates as they impart ``maintenance mode'' feedback onto their hosts \citep[e.g.,][]{Best2006,Kormendy2013}.  

A number of studies consider the cosmic spin evolution of massive BHs based on models in which corotating accretion flows always spin a BH up \citep[e.g.,][]{King2008,Barausse2012,Volonteri2013,Izquierdo-Villalba2020}.  We speculate that reversing this assumption for thick disks may significantly reduce the spins of supermassive BHs in the most massive galaxies, as well as super-Eddington accretors. This may help models reproduce the observed population of billion solar mass quasars at $z\sim 6-7$ \citep[e.g.,][]{Shapiro2005,Volonteri2005,Zubovas2019}.  As a reminder, our results pertain specifically to hot accretion flows in the MAD state. The spin-down effects we describe will be much less severe in the opposite SANE case. The results described in \citet{Gammie2004} may be relevant in that limit. In future work, we plan to explore the spin evolution of super-Eddington disks in more detail.

In the case of M87*, the current mass accretion rate is estimated to be $\dot{M}_0 \sim 10^{-3}M_\odot{\rm yr^{-1}}$ \citep{EHTVIII}, which corresponds to $f_{\rm Edd} \gtrsim 10^{-5}$. At this mass accretion rate, the BH spin will remain essentially unchanged even over a time as long as the age of the universe. However, it appears that in the past, M87* had a much more powerful jet, with power reaching perhaps $10^{45} {\rm erg\,s^{-1}}$ \citep{Owen2000,deGasperin2012}, which is $\sim 100$ times greater than the current jet power. If that power level had been maintained for a Gyr, then M87* would have experienced significant spindown.

\subsection{Generality of the Results}

It is important to keep in mind that the results presented in this paper apply only to radiatively inefficient hot accretion flows. Accretion systems where radiative cooling is important are quite different, and our results do not apply to those. Even within the class of radiatively inefficient hot flows, our work focuses only on the MAD regime. The opposite case of SANE accretion, where $\phi_{\rm BH}$ lies below the saturation value $\phi_{\rm BH,sat}$, needs to be explored separately. An added complication in SANE accretion is that, in addition to the BH spin $a_*$, the results will depend also on a second parameter, viz., the amount of magnetic flux at the BH horizon relative to the saturation value: $\phi_{\rm BH} / \phi_{\rm BH,sat}$.

As discussed in Section~\ref{sec:intro}, there is some observational evidence that the MAD regime may be reasonably common in Nature. There is also theoretical evidence that this regime is easier to achieve in long-lived systems than previously thought. (In this context, almost any system in Nature is extremely long-lived compared to the time scales probed by simulations.) The good agreement between our results and those reported in \citet{Tchekhovskoy2012}, even though the two sets of simulations differ by a factor of 3 in duration, gives confidence that the results are reasonably well-converged, and may be applied to MAD systems. 

There is, however, a caveat. Once $\phi_{\rm BH}$ at the horizon has reached the MAD saturation limit $\phi_{\rm BH,sat}$, flux accumulates in the surrounding accretion flow and we expect a MAD-like ``magnetosphere" to develop out to some radius $r_{\rm MAD}$ in the disk \citet{Avara_2016} discuss a similar idea in the context of a geometrically thin disc model.
If enough magnetic flux of the same sign (no reversals in $B_z$) is supplied by the accretion flow, $r_{\rm MAD}$ will increase monotonically with time. For the simulations described in this paper, we think $r_{\rm MAD}$ is of order several tens of $r_g$ (perhaps as much as $100r_g$)\footnote{The models in this paper have achieved inflow equilibrium out to radii well in excess of $100r_g$. This is one of the benefits of running long-duration simulations.}, though we do not have a reliable method of defining $r_{\rm MAD}$. The change in the character of the radial profiles of $\rho$ and $u^r$ in Figure \ref{fig:radial} at $r\approx 50r_g$ might suggest that this radius corresponds to the location of $r_{\rm MAD}$ (we thank the referee for this suggestion). We imagine that $r_{\rm MAD}$ was a little smaller for the simulations in \citet{Tchekhovskoy2012}, but perhaps not by a large factor. On the other hand, the stellar-winds-driven accretion model of Sgr A$^*$ described by \citet{ressler_2020a,ressler_2020b} conceivably had $r_{\rm MAD}$ as large as $10^4r_g$. This brings up the following question: Could the properties of MADs change substantially if $r_{\rm MAD}$ is very much larger than the typical values explored so far via simulations? If the answer is yes, then $r_{\rm MAD}$ would become a relevant second parameter (in addition to $a_*$) in the MAD regime of accretion, and its effects will need to be quantified.

\section{Summary}
\label{sec:summary}
\begin{table*}
\centering
\renewcommand{\arraystretch}{1.3}
\begin{tabular}{c | c c c c c c c c c}
\hline\hline
\vspace*{0mm}

Model & $\phi_{\rm BH}$ & $\eta$ & $s$ & $s'$ & $\sigma/\mu$ & $\sigma/\mu$ & $h/r$ & $w_{\rm jet}$ & $k$\\
BH spin $a_*$ &&&&&$(\dot{M}_0)$&$(\phi_{\rm BH})$&$(r=3\,r_{\rm g})$& $(z=10\,r_{\rm g})$ & \\

\hline
0.9 & 56.2 & 1.31 & $-8.88$ & $-6.79$ & 0.342 & 0.209 & 0.144 & 18.2 & $0.428$\\
0.7 & 62.1 & 0.711 & $-8.30$ & $-11.7$ & 0.309 & 0.176& 0.119 & 20.4 &$0.407$\\
0.5 & 63.6 & 0.343 & $-6.18$ & $-18.1$ & 0.260 &0.160 & 0.107 & 20.0 &$0.385$\\
0.3 & 62.3 & 0.140 & $-3.87$ & $-27.7$ & 0.216 & 0.122& 0.106 & 21.3 & $0.370$\\
0 & 50.0 & 0.0345 & 0.485 & - & 0.156 & 0.094& 0.173& - & - \\
$-0.3$ & 42.2 & 0.0718 & 3.35 & $-46.7$ & 0.163 & 0.104 &0.207 & 17.5 & $0.273$\\
$-0.5$ & 35.6 & 0.104 & 3.99 & $-38.2$ & 0.167 &0.129 &0.242 & 14.5 & $0.351$\\
$-0.7$ & 27.9 & 0.126 & 4.25 & $-33.8$ & 0.149 & 0.158&0.300 & 11.9 & $0.355$\\
$-0.9$ & 24.5 & 0.187 & 4.65 & $-24.8$ & 0.179 & 0.176 & 0.315& 11.6 & $0.418$\\

\hline

\end{tabular}
\caption{Summary table of simulation results. For each of the nine simulations we provide the magnetic flux parameter $\phi_{\rm BH}$, the jet efficiency $\eta$, the spinup parameter $s$, the modified spinup parameter $s'$, the variability $\sigma/\mu$ in both the accretion rate $\dot{M}_0$ and $\phi_{\rm BH}$, the disk scale height $h/r$ at radius $3\,r_{\rm g}$, the jet width at height $z=10\,r_{\rm g}$, and the best-fit power-law index $k$ for the average jet shape. All quantities were computed from time- and azimuth-averaged data over time range $50,000\,t_{\rm g}$ to $100,000\,t_{\rm g}.$} 
\label{tab:summary}
\end{table*}
In this work, we explored the long time evolution of radiatively-inefficient magnetically arrested disks (MADs) for nine different values of the black hole (BH) spin parameter, $a_* = -0.9$, $-0.7$, $-0.5$, $-0.3$, 0, 0.3, 0.5, 0.7 and 0.9, using the GRMHD code KORAL. We evolved our simulations up to $t \gtrsim 10^5t_g$ to ensure inflow equilibrium out to large radii. We considered the effect of BH spin on the dimensionless magnetic flux parameter $\phi_{\rm BH}$ and the jet efficiency $\eta$, and found results in agreement with previous work by \citet[][]{Tchekhovskoy2012} which used shorter duration simulations.  We also estimated the spindown rate in MAD geometrically thick accretion flows.

In Table~\ref{tab:summary}, we present a summary of our time-averaged results in the nine simulations for the following quantities: the dimensionless magnetic flux on the horizon, the jet efficiency, the regular and modified spinup parameters, variability in the mass accretion rate and the magnetic flux parameter, disk scale height at $r=3r_g$, jet width at $z=10r_g$, and power-law index $k$ of the time-averaged jet shape. 

\medskip\noindent Our main conclusions are:
\begin{itemize}
    \item The saturation value of the magnetic flux of MAD disks depends on the BH spin. Retrograde disks saturate at a lower relative magnetic flux than prograde systems. 
    \item Prograde MAD systems produce more powerful jets than retrograde systems. The jet is powered by the BH spin energy in all cases, but the lower magnetic flux saturation level in retrograde systems limits their jet power and efficiency. 
    \item All jets exhibit a parabolic shape with a power-law index of $k \approx 0.27-0.42$, similar to values observed in AGN jets.
    \item Retrograde MAD simulations have narrower jets and thicker equatorial disks near the BH, compared to prograde systems with the same spin magnitude. Thus, given a BH where there is evidence that it is a MAD system (as in the case of M87*, \citealt{EHTVIII}), one could potentially constrain the dimensionless magnetic flux $\phi_{\rm BH}$ and the BH spin $a_*$ using a measured jet width or disk scale height close to the BH.
    \item Prograde and retrograde MADs exhibit different variability trends in accretion rate, with variability increasing with increasing spin for $a_*>0$, while remaining almost constant for $a_*<0$. Variability in the magnetic flux on the BH shows the opposite behavior.
    \item At all nonzero spins, jets from MAD systems spin down the BH by sapping it of angular momentum. If jets are persistent over cosmic time, this spindown can notably reduce the BH spin.  
\end{itemize}

We have neglected the effect of radiation in the simulations described in this work (although KORAL is equipped to include radiation when needed), and hence, our target BH systems are low-luminosity AGN and low-hard state BH binaries, where the accreting gas is radiatively inefficient and the accretion disk is geometrically thick. In future studies, it will be important to understand how our results will change if we consider radiatively-efficient thin disks as well super-Eddington radiatively-supported thick accretion flows \citep[e.g.,][]{Sadowski2014}. Further, it would be interesting to understand whether the $\phi_{\rm BH}$ vs $a_*$ relationship would change under more general disk geometries with a misalignment between the spin vectors of the BH and the disk \citep{Fragile2007,Liska2018}. Recent work suggests that the magnetic flux onto the BH drops with higher misalignment angles given the same initial disk magnetic field \citep[e.g.,][]{Chatterjee2020}.  

\section{acknowledgements}

We thank Jason Dexter, Razieh Emami-Meibody and Oliver Porth for helpful comments. This work was supported in part by the National Science Foudation under Grants OISE-1743747 and AST1816420. The simulations were performed using high-performance computing resources on Frontera under Frontera Large-Scale Community Partnership (LSCP) allocation AST20023, on Stampede2 under Extreme Science and Engineering Discovery Environment (XSEDE) allocation AST080028, and on computer clusters at the Black Hole Initiative (BHI). Computations at the BHI were made possible through the support of grants from the Gordon and Betty Moore Foundation and the John Templeton Foundation. The opinions expressed in this publication are those of the authors and do not necessarily reflect the views of the Moore or Templeton Foundations. 
AC is supported by Hubble Fellowship grant HST-HF2-51431.001-A awarded by the Space Telescope Science Institute, which is operated by the Association of Universities for Research in Astronomy, Inc., for NASA, under contract NAS5-26555.

\section{Data Availability}

The data used in the work presented in this article is available upon request to the corresponding author.


\bibliography{spindown.bib}{}
\bibliographystyle{mnras}

\appendix

\section{Effect of the adiabatic index}
\label{sec:adiab}

\begin{figure*}
	\includegraphics[width=0.9\textwidth]{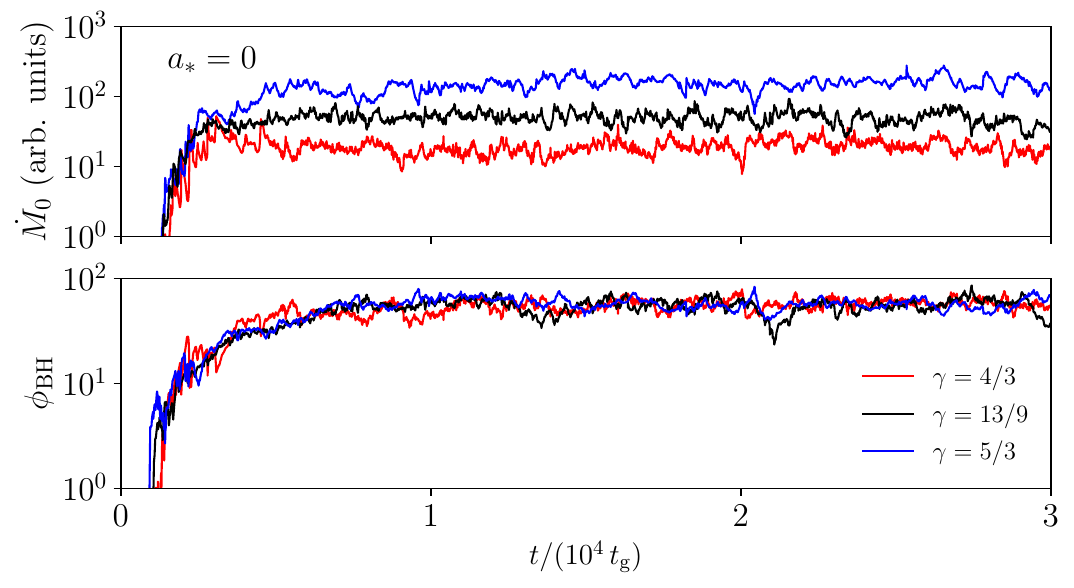}
    \caption{We compare the accretion rate $\dot{M}_0$ (top row) and the dimensionless magnetic flux parameter $\phi_{\rm BH}$ (bottom row) at the BH horizon for simulations with BH spin $a_*=0$ and three different choices of the adiabatic index: $\Gamma=4/3$ (red), $\Gamma=13/9$ (the fiducial model in the main text, black), and $\Gamma=5/3$ (blue). The mean accretion rate (in arbitrary units) is sensitive to the choice of adiabatic index, but the dimensionless magnetic flux $\phi_{\rm BH}$ saturates at the same value in all three simulations.
    }
    \label{fig:compare_Gamma}
\end{figure*}

We have checked whether our choice of adiabatic index $\Gamma=13/9$ has a significant impact on the magnetic flux accumulated on the black hole in our simulations. We ran two additional simulations, one with $\Gamma=4/3$ and the other with $\Gamma=5/3$, for BH spin $a_*=0 $. Both simulations were run up to a total time of $t=30,000\,t_{\rm g}$. The radius of the pressure maximum of the initial torus needed to be adjusted in these simulations in order to keep the outer edge of the torus at $r\approx10^4\,r_{\rm g}$; for $\Gamma=4/3$ we set $r_{\rm max}=42.43r_{\rm g}$, for $\Gamma=5/3$ we set $r_{\rm max}=42.40r_{\rm g}$, while for our fiducial $\Gamma=13/9$ we set $r_{\rm max}=42.43$ (\autoref{tab:init}). The simulation grid and all other initial conditions were the same as in our fiducial simulation.

\autoref{fig:compare_Gamma} shows the accretion rate $\dot{M}_0$ and  dimensionless magnetic flux  $\phi_{\rm BH}$ at the horizon for the three $a_*=0$ simulations. We find that the accretion rate in arbitrary units increases with increasing adiabatic index, but this is of no consequence since all our results correspond to dimensionless quantities for which the BH mass and mass accretion rate are scaled out. 

The dimensionless magnetic flux parameter $\phi_{\rm BH}$ is plotted in \autoref{fig:compare_Gamma} as a function of time for the three simulations. This quantity shows no dependence on the adiabatic index. In particular, $\phi_{\rm BH}$ saturates at essentially the same value, $\phi_{\rm BH}\approx 50$, in all three simulations. Variability and other dimensionless diagnostics are also similar for different values of $\Gamma$.

\section{Checking the MAD saturation level for retrograde spins}
\label{sec:MAD_phibh}
\begin{figure*}
	\includegraphics[width=0.9\textwidth]{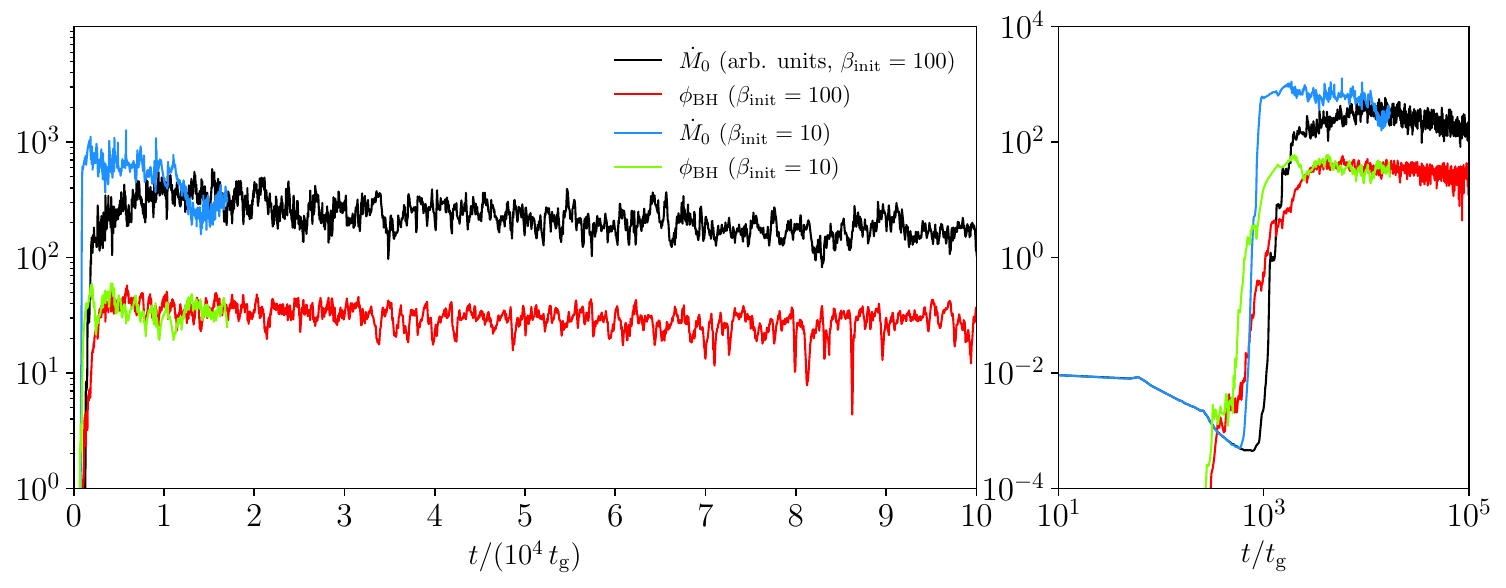}
    \caption{We compare the accretion rate $\dot{M}_0$ (black line) and the dimensionless magnetic flux parameter $\phi_{\rm BH}$ (red line) at the BH horizon from a spin $a_*=-0.7$ simulation run with our fiducial initial magnetic field strength ($\beta_{\rm init}=100$) and from a simulation with an initial magnetic pressure 10 times larger ($\beta_{\rm init}=10$; blue and green lines). (Right) the same data plotted on a log-log scale to emphasize the differences between the simulations in the initial, transitory phase, before accretion reaches steady-state. Despite the differences in the initial conditions, both simulations saturate at the same dimensionless magnetic flux $\phi_{\rm BH}$. 
    }
    \label{fig:compare_phibh}
\end{figure*}

Figure~\ref{fig:phi_eta_MAD} shows that our retrograde simulations saturate at a significantly lower value of the dimensionless magnetic flux parameter $\phi_{\rm BH}$ than the prograde simulations. In order to check whether our retrograde spin cases have reached their maximum value of magnetic flux and to test the impact of our initial conditions, we ran an additional simulation with the same parameters and initial conditions as for spin $a_*=-0.7$, but with an initial magnetic field strength $|b|$ stronger by a factor of $\sqrt{10}$ (i.e., initial $\beta_{\rm init}$ lower by a factor of 10). The simulation grid and all other initial conditions were identical. We ran this new simulation up to a total time of $t=17,000\,t_{\rm g}$. 

Figure \ref{fig:compare_phibh} shows that, in the initial, transitory period $t \lesssim 2000\,t_{\rm g}$, the simulation with the stronger initial field $(\beta_{\rm init}=10)$ does have a higher value of $\phi_{\rm BH}$ than our fiducial setup $(\beta_{\rm init}=100)$. However, once the simulations reach steady-state and are fully accreting gas ($t\gtrsim2000\,t_{\rm g}$), both simulations saturate at the same mean value of $\phi_{\rm BH}\approx30$.  This test suggests that the saturation levels of the MAD simulations run in this paper do not depend on the initial field strength in the torus.

\end{document}